\documentclass[aps,prb,letter,twocolumn,showpacs,amsfonts,amssymb,amsmath]{revtex4-2}

\usepackage{graphicx}
\usepackage{natbib}
\usepackage{bm}
\usepackage{bbm}
\usepackage{physics}
\usepackage{amsmath}
\usepackage{amsthm}
\usepackage{dsfont}
\usepackage{dcolumn}
\usepackage[colorlinks,citecolor=blue,linkcolor=blue,urlcolor=blue,bookmarks=false,hypertexnames=true]{hyperref}
\usepackage[mathscr]{eucal}

\begin{document}

\title{Decoherence and fidelity enhancement during shuttling of entangled spin qubits}
\author{Yu-Ning Zhang}
\thanks{These two authors contributed equally}
\author{Aleksandr S. Mokeev}
\thanks{These two authors contributed equally}
\author{Viatcheslav V. Dobrovitski}
 \email{V.V.Dobrovitski@tudelft.nl}
 \affiliation{QuTech and Kavli Institute of Nanoscience, Delft University of Technology,\\
 PO Box 5046, 2600 GA Delft, The Netherlands}

\date{\today}

\begin{abstract}
Shuttling of spin qubits between different locations is a key element in many prospective semiconductor systems for quantum information processing, but the shuttled qubits should be protected from decoherence created by time- and space-dependent noises. Since the paths of different spin qubits are interrelated, 
the noises acting on the shuttled spins exhibit complex and unusual correlations. 
We appraise the role of these correlations using the concept of trajectories on random sheets, and demonstrate that they can drastically affect efficiency of the coherence protection. These correlations can also be exploited to enhance the shuttling fidelity, and we show that by encoding logical qubit in a state of two consequtively shuttled entangled spins, high fidelity can be achieved even for very slow shuttling. We identify the conditions favoring this encoding, and quantify improvement in the shuttling fidelity in comparison with the single-spin shuttling.
\end{abstract}

\maketitle

Shuttling of spin qubits between different spatial locations is an important component of prospective scalable semiconductor platforms for quantum information processing \cite{FaultTolQuDotTaylorLukin05,ShuttlingSAWMeunier11,BenjaminBaughEtalTopoQCSi19,VdSBluhmVeldhorstEtalVision17,langrock_blueprint_2023,KunneEtalSpinBusArch23,BauerleTakadaEtalAcousticTransportElectrons22,
mills_shuttling_2019,
BenjaminEtalMulticoreQComp22,BoterVdsEtalSpiderWebArray22,noiri_shuttling-based_2022,WangVeldhorstEtalGeQDs24}. Within this approach, an electron (or a hole) in a semiconductor structure is adiabatically carried along the shuttling channel by an external time-varying potential, created e.g.\ by a set of gates \cite{seidler_conveyor-mode_2022,langrock_blueprint_2023,StruckSchreiberEtalSpinPairShuttling23,noiri_shuttling-based_2022,SmetVandersypenEtal24SpinShuttlSilicon,WangVeldhorstEtalGeQDs24,yoneda_coherent_2021,VolmerStruckEtalValleySplit23} or by a surface acoustic wave (SAW)
\cite{ShuttlingSAWMeunier11,
JadotMeunierEtalTwoSpinShuttling21,BauerleTakadaEtalAcousticTransportElectrons22,ShuttlingSAWRitchie11,
HuangHuSpinRelaxShuttl13
}. 
The electron propagates in the form of a wavepacket tightly localized in space, transporting the spin qubit from one quantum dot to another. 
During the shuttling process, time- and space-dependent magnetic noise dephases the electron spin \cite{seidler_conveyor-mode_2022,yoneda_coherent_2021,
van_riggelen-doelman_coherent_2023,
noiri_shuttling-based_2022,StruckSchreiberEtalSpinPairShuttling23, SmetVandersypenEtal24SpinShuttlSilicon,JadotMeunierEtalTwoSpinShuttling21,BoterJoyntVdSNoiseCorrBellStates20,MortemousqueMeunierEtalShuttl2DArray21,WangVeldhorstEtalGeQDs24}, such that 
protecting coherence of the shuttled qubit is critical for achieving fault tolerance. Increasing the shuttling speed has limited potential to suppress decoherence, because fast shuttling gives rise to many other decoherence mechanisms due to the loss of adiabaticity, excitation of the carrier to higher orbital and valley states in the semiconductor, and additional decoherence due to spin-orbit coupling \cite{langrock_blueprint_2023,HuangHuSpinRelaxShuttl13,BoscoZouLossHighFidShuttlingSOI23,VolmerStruckEtalValleySplit23,LosertFriesenValleySplit23,HaoRuskovTahanEtal14}. Therefore, exploration of alternative coherence protection approaches and their realistic assessment are timely and important tasks. 

In this work we consider a paradigmatic method of coherence protection, when the state of a logical qubit is encoded in a singlet-triplet (ST) decoherence-free subspace of two spins \cite{ZanardiRasettiDFS97,
ViolaCoryEtalDFSExp01,BurkardLaddPanNicholReviewQuDots23}, which are consequtively shuttled one after another. Feasibility of this approach has been demonstrated \cite{JadotMeunierEtalTwoSpinShuttling21}, but the fidelity enhancement in various regimes has not been analyzed. 
Here we show, how the spatiotemporal correlations of the noise play crucial role in decoherence of ST qubits. We calculate the shuttling fidelity, and identify the range of parameters where the ST encoding substantially enhances  fidelity. Importantly, we show that shuttling with arbitrarily high fidelity over arbitrarily long distances is possible even for very low shuttling speed, making ST encoding a promising way to reach the fault tolerance threshold in the long-range shuttling architectures.

\begin{figure}[tbp]
\centering
\includegraphics[width=\columnwidth]{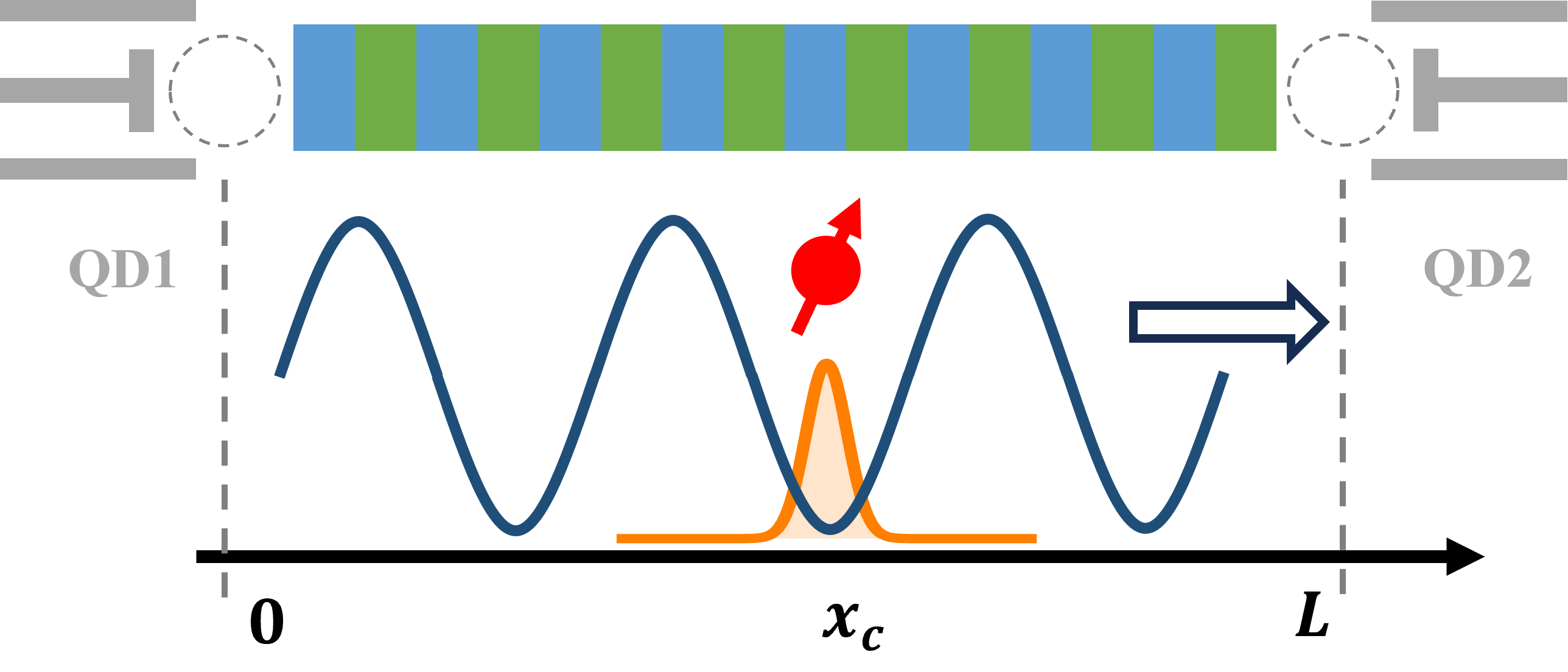}
\caption{Top: schematic representation of the spin shuttling from quantum dot QD1 (left) to QD2 (right), using the running-wave confining potential generated by the clavier gates (blue/green rectangles). Bottom: the electron, confined within a minimum of the moving potential (solid dark blue line), propagates as a tightly localized wavepacked (orange) with the center at $x=x_c$. The electron spin (red arrow) encodes the qubit state.}
\label{fig:shuttling}
\end{figure}

We consider conveyor-belt type shuttling, where  a long (length $L\sim 1$--10~$\mu$m) one-dimensional channel is formed between two quantum dots, see Fig.~\ref{fig:shuttling}. 
The shuttled electron, confined in a moving potential well, propagates as a tightly localized wavepacket with the center at the point $x=x_c$ and the spatial distribution of the electron density $\rho(x,y,z;x_c)\equiv\rho({\vec r};x_c)$. The wavepacket is shuttled adiabatically, staying within the minimum of the shuttling potential without leaking to the adjacent minima or excitation to higher orbital states or to the excited valley states. 
The moving confining potential can be created in Si- or Si/SiGe systems 
using a set of clavier gates deposited on top of the shuttling channel
\cite{seidler_conveyor-mode_2022,langrock_blueprint_2023,
StruckSchreiberEtalSpinPairShuttling23}.
In GaAs-based structures, a SAW propagating along the shuttling channel creates the moving confining potential via piezoelectric effect 
\cite{ShuttlingSAWMeunier11,ShuttlingSAWRitchie11,HuangHuSpinRelaxShuttl13,
JadotMeunierEtalTwoSpinShuttling21, BauerleTakadaEtalAcousticTransportElectrons22}.

Ideally, the shape of the wavepacket would stay constant during shuttling, with  $\rho({\vec r};x_c)=\rho_0(x-x_c,y,z)$, but in reality it fluctuates in time and space, because the 
random variations of the material properties along the shuttling channel, as well as defects which randomly trap/release charges at different locations (charge traps), affect the confining potential and lead to small random displacements and distortions of the wavepacket
\cite{BoscoZouLossHighFidShuttlingSOI23,HuangHuSpinRelaxShuttl13,StruckCywinskiSchreiber20QubitNoise,KepaCywinskiKrzywdaSpinNoise23,zou_spatially_2023,ShalakDelerueNiquetChargeNoiseSiHole23,SpenceNiquetMeunierEtalChargeNoise22,ShehataVanDorpeEtalChargeNoiseQuDots23,BurkardLaddPanNicholReviewQuDots23}. 
%
%
In a typical experiment, a nominally uniform quantizing magnetic field $B_Q\sim 0.1$--1~T is applied to the system, but because of the random variations in the material properties, the $g$-factor of the electron slightly varies in space, such that the field experienced by the electron depends on its position. Therefore, random variations of the electron wavepacket in space and time lead to corresponding random variations of the Zeeman energy of the electron spin 
\footnote{
We assume that the {\em direction} of the effective magnetic field is almost uniform in space and time, such that the longitudinal relaxation can be neglected; this is a good approximation for Si-based structures. In GaAs systems the longitudinal relaxation due to spin-orbit coupling is important \cite{HuangHuSpinRelaxShuttl13}, while for Ge-based systems the spatial variation of the direction of the effective magnetic field should be taken into account \cite{van_riggelen-doelman_coherent_2023, WangVeldhorstEtalGeQDs24,BoscoZouLossHighFidShuttlingSOI23}, and the approach presented here needs some modifications. 
}.
Besides, the randomly located nuclear spins create a position-dependent hyperfine field $B_{\rm hf}({\vec r},t)$, which fluctuates in space and time (e.g.\ due to nuclear spin flip-flops).
Thus, the spin $S=1/2$ of an electronic wavepacket with the center at $x_c$ experiences the Zeeman splitting
\begin{equation}
\label{eq:tildeB}
{\tilde H}_Z (x_c,t) = g_0 \mu_B \left[B_0(x_c) + {\tilde B}(x_c,t)\right] S_z, 
\end{equation}
see \cite{SupplementalMaterial} for details,
where $g_0$ is a nominal $g$-factor of the electron in the shuttling channel (the appropriately averaged value of $g({\vec r})$), while $B_0(x_c)$ and ${\tilde B}(x_c,t)$ represent, respectively, the deterministic and the random part of the effective magnetic field acting on the shuttled spin. For simplicity, everywhere below we set $g_0\mu_B=1$.

An electron spin shuttled along the spacetime trajectory $x_c(t)$, starting at $x_c=0$ at $t=0$ and finishing at $x_c=L$ at $t=t_f$, experiences a time-dependent magnetic field comprised of the deterministic $B_0(t)\equiv B_0(x_c(t))$ and the random ${\mathrm B}(t)\equiv {\tilde B}(x_c(t),t)$ parts. During shuttling, the spin acquires the phase $\Theta=\alpha+\phi$, where
\begin{equation}
\alpha = \int_0^{t_f} B_0(t)\,dt\ \ {\rm and}\ \ 
\phi = \int_0^{t_f} {\mathrm B}(t)\,dt.
\end{equation}
The deterministic phase $\alpha$ can be taken into account during post-processing, and is set to zero everywhere below. In contrast, the random phase $\phi$ leads to the loss of fidelity $F$, which is quantified by the dephasing factor $W$ or the dephasing exponent $\chi$ as 
\begin{equation}
\label{eq:wRS}
W=\exp{-\chi} = {\mathbb E}\,\exp{-i\phi},
\end{equation}
where the expectation ${\mathbb E}$ corresponds to averaging over the random process ${\mathrm B}(t)$, and the shuttling fidelity is $F=(1+W)/2$.

Statistical properties of the process ${\mathrm B}(t)$ are determined by the trajectory $x_c(t)$ and by the underlying space- and time-dependent noise ${\tilde B}(x_c,t)$. 
In this work we treat the noise ${\tilde B}(x_c,t)$ using the mathematical concept of random sheet \cite{Chentsov_1956,Kitagava_1951,AdlerBookGeomRandF}, which generalizes the idea of a random process. We model ${\tilde B}(x_c,t)$ as a Gaussian random sheet, which is defined by its mean, which we set to zero (non-zero mean can be included in the deterministic phase $\alpha$), and the two-point covariance function $K(x_c,t;x_c',t')$. 
The model of a Gaussian random sheet presents the same advantages \cite{MokeevZhangDobrovitski24} as the model of a Gaussian random process in the theory of dephasing of stationary qubits \cite{Kubo1954,KlauderAnderson62,
ChandrasekharRandomProc}. A Gaussian sheet can be viewed as a sum of noise fields created by a large number of weak independent sources. Gaussian sheet realistically describes the finite span of the noise correlations in time and space, and at the same time is easily tractable by analytical or numerical means. Namely, if ${\tilde B}(x_c,t)$ is a Gaussian sheet, then for any realistic spin trajectory $x_c(t)$, the corresponding random process ${\mathrm B}(t)$ is a Gaussian random process with the covariance function $K_{\mathrm B}(t,t')=K(x_c(t),t;x_c'(t'),t')$. Then the averaging in Eq.~(\ref{eq:wRS}) can be performed analytically, yielding 
\begin{equation}
\chi=\frac{1}{2} \int_{0}^{t_f} \int_{0}^{t_f}\! K_{\mathrm{B}}(t,s)\, dt\, ds,
\end{equation}
if the trajectory $x_c(t)$ is continuous.

\begin{figure}[tp]
\includegraphics[width=\linewidth]{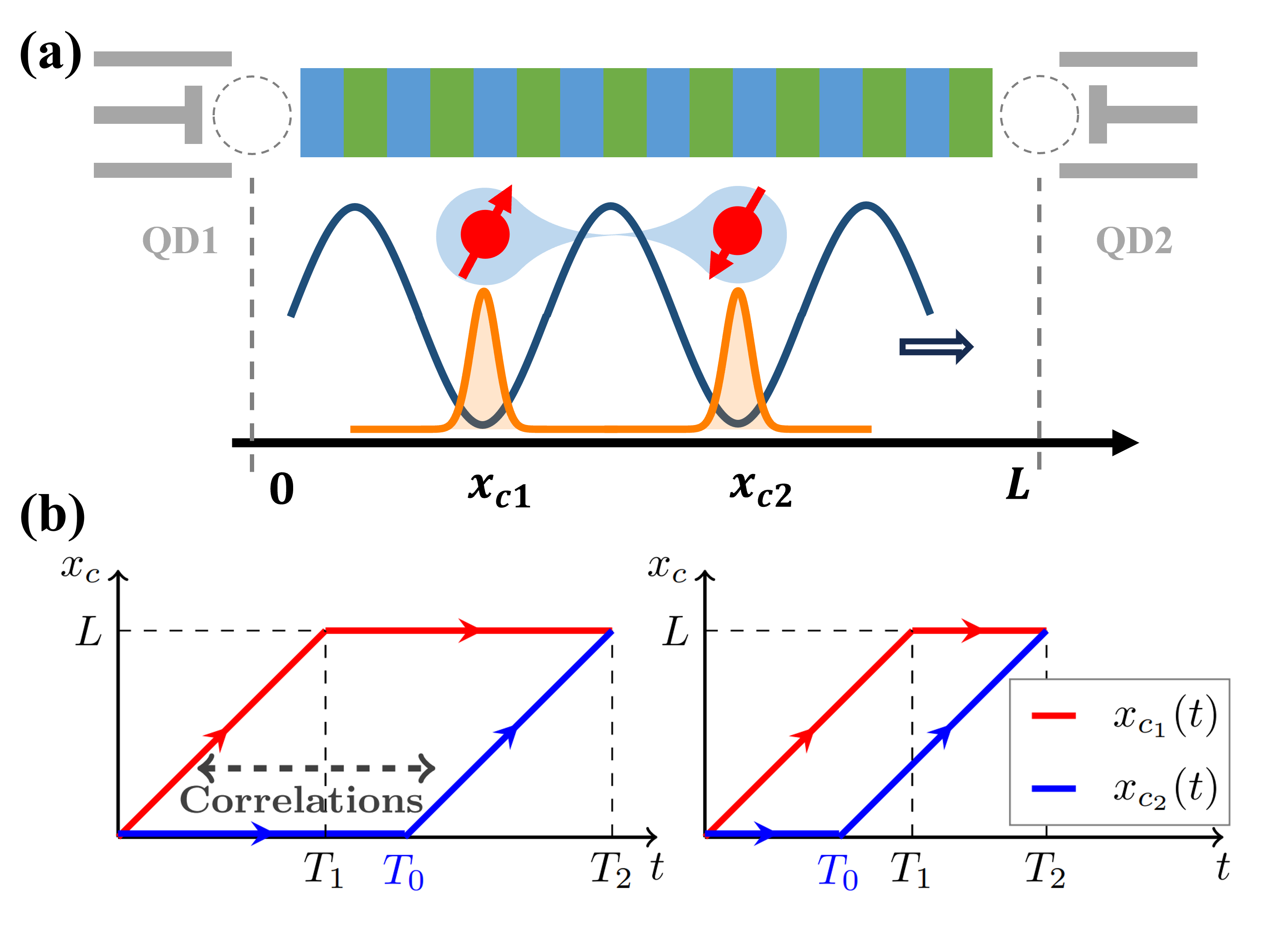}
\caption{{\bf (a)}: Sequential two-spin shuttling: the first electron is loaded into the shuttling channel at $t=0$ and is shuttled with the constant velocity $v$. After the delay $T_0$, the second electron is loaded into the channel and is shuttled with the same velocity. During shuttling, both electrons are confined in the minima of the moving potential (solid dark blue line), forming localized wavepackets (orange) with centers at $x=x_{c1}$ and $x=x_{c2}$.
After the first electron reaches the destination $x_{c1}=L$ at $t=T_1$, it stays there while the second electron arrives to $x_{c2}=L$ at $t=T_2$, such that the total shuttling time $t_f=T_2$.
{\bf (b)}: Space-time trajectories of the shuttled spins, $x_{c1}(t)$ (red solid line) and $x_{c2}(t)$ (blue solid line). Depending on the travel time $T_1$ and the delay $T_0$, two cases are possible, $T_0>T_1$ and $T_0<T_1$, shown in left and right panels, respectively.}
\label{fig:two_spin_sequence}
\end{figure}

A particularly illustrative model is Ornstein-Uhlenbeck (OU) sheet with the covariance function $K_{\mathrm{OU}}(x_c,t;x_c',t') = \sigma_{\mathrm B}^2\,\exp{-\kappa_x |x_c-x_c'| - \kappa_t |t-t'|}$, describing the correlations exponentially decaying in space and in time \cite{langrock_blueprint_2023,rojas-arias_spatial_2023,zou_spatially_2023,yoneda_noise-correlation_2023,StruckCywinskiSchreiber20QubitNoise,CywinskiWitzelDasSarmaDD09,DobrEtal09OUnoise}, with the correlation length $\lambda_c=1/\kappa_x$ and the correlation time $\tau_c=1/\kappa_t$. If the qubit is encoded in the spin of a single electron, shuttled from $x_c=0$ to $x_c=L$ with the constant velocity $v$ (such that $x_c(t)=vt$), then the noise acting on the qubit is an OU process with the covariance 
$K_{\mathrm B}(t,s)= \sigma_{\mathrm B}^2 \exp{-\kappa|t - s|}$,
where $\kappa = \kappa_x v + \kappa_t$, and the dephasing factor $W_1=\exp{-\chi_1}$ for the single-spin shuttling is 
\begin{equation}
\label{eq:wT}
\chi_1=(\sigma_{\mathrm B}/\kappa)^2 \,[\kappa t_f+\exp{-\kappa t_f}-1],
\end{equation}
where $t_f=L/v$. 
In order to minimize the dephasing, the shuttling velocity should be increased. However, shuttling at high speed leads to the loss of adiabaticity, leakage to the neighboring minima, and excitation to higher orbital and valley states, as well as rapid spin relaxation due to spin-orbit coupling \cite{langrock_blueprint_2023,HuangHuSpinRelaxShuttl13,BoscoZouLossHighFidShuttlingSOI23,VolmerStruckEtalValleySplit23,LosertFriesenValleySplit23}.

\begin{figure}[tbp!]
\includegraphics[width=1\columnwidth]{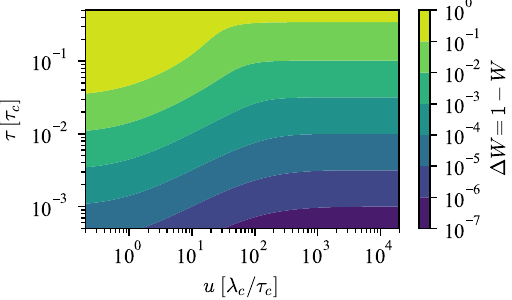}
\caption{Fidelity loss $\Delta W$ during shuttling of two entangled spins as a function of the dimensionless delay time $\tau=\kappa_t T_0=T_0/\tau_c$ and dimensionless velocity $u=v\,\tau_c/\lambda_c=v\,\kappa_x/\kappa_t$. The spins are shuttled over the distance $L=100\,\lambda_c$, the noise amplitude is $\sigma_{\mathrm B}=\sqrt{2}\,\kappa_t$. With the experimental parameters given in Table~\ref{tab:experimental_params}, this corresponds to $L = 10\,\mu\textrm{m}$, $\tau_c=20\,\mu\textrm{s}$, $\lambda_c=0.1\,\mu \textrm{m}$, and the dephasing time of a static spin $T_{2,s}^*= 20\,\mu\textrm{s}$. The dimensional shuttling velocity $v$ would be measured in units of $\lambda_c/\tau_c=5\,\mathrm{mm}/{\mathrm s}$.}
\label{fig:F_surface_scan}
\end{figure}

An alternative way to enhance the shuttling fidelity is to exploit the correlations of the noise, encoding the state of the logical qubit $a|0\rangle + b|1\rangle$ in a decoherence-free subspace \cite{ZanardiRasettiDFS97,
ViolaCoryEtalDFSExp01} formed by the singlet $|\Psi^-\rangle$ and the triplet $|\Psi^+\rangle$ states 
of two electron spins, as $|\psi_0\rangle = a\,|\!\uparrow_1\downarrow_2\rangle + b\,|\!\downarrow_1\uparrow_2\rangle = a'\,|\Psi^+\rangle + b'\,|\Psi^-\rangle$,
where $|\Psi^\pm\rangle = \left(|\!\uparrow_1\downarrow_2\rangle \pm |\!\downarrow_1\uparrow_2\rangle\right)/\sqrt{2}$. The two spins are shuttled through the same channel with the delay $T_0$ one after another, see Fig.~\ref{fig:two_spin_sequence}.
The spins shuttled along the trajectories $x_{c1}(t)$ and $x_{c2}(t)$ will experience different but correlated noises ${\mathrm B}_1(t)={\tilde B}(x_{c1}(t),t)$ and ${\mathrm B}_2(t)={\tilde B}(x_{c2}(t),t)$ produced by the same random sheet ${\tilde B}(x_c,t)$. The random phases acquired by the spins transform the state of the logical qubit into $|\psi_\phi\rangle = a\,|\!\uparrow_1\downarrow_2\rangle + b\exp{i\phi_2}\,|\!\downarrow_1\uparrow_2\rangle$, and the shuttling fidelity is determined by the dephasing factor $W_2 = \mathbb{E}\,e^{i\phi_2}$ with
\begin{equation}
\label{eq:wtwospins}
\phi_2 = \int_0^{t_f} \left({\mathrm B}_1(t)-{\mathrm B}_2(t)\right)\,dt, 
\end{equation}
where the total shuttling time $t_f=T_2$, see Fig.~\ref{fig:two_spin_sequence}.
If the noise ${\tilde B}(x_c,t)$ varies slowly then ${\mathrm B}_1(t)$ and ${\mathrm B}_2(t)$ are correlated and the random phase $\phi_2$ is small, thus increasing the fidelity.

\begin{table}[tbp!]
\caption{Typical parameters of a realistic device.}
\label{tab:experimental_params}
\begin{ruledtabular}
\begin{tabular}{lll}
Symbol  & Range  & Description
\\ 
\hline
$v$  & $10^{-3}$--$10^2\,\textrm{m/s}$  & Shuttling velocity \\ 
$L$     & $10\,\mu \textrm{m}$  & Shuttling length   \\ 
$\tau_{c}$  & $20\,\mu \textrm{s}$ & Correlation time of $\tilde{B}(x,t)$ \\
$\lambda_{c}$  & $100\, \textrm{nm}$ & Correlation length of $\tilde{B}(x,t)$ \\
$T_{2,s}^{*}$ & $20\,\mu \textrm{s}$ & Dephasing time of a static spin
\end{tabular}
\end{ruledtabular}
\end{table}

In order to quantify the fidelity enhancement and identify the favorable region of parameters, we calculate the two-spin dephasing factor $W_2=\exp{-\chi_2}$ for OU random sheet ${\tilde B}(x_c,t)$; an explicit analytical answer can be obtained \cite{SupplementalMaterial} due to Gaussian nature of the noise. Fig.~\ref{fig:F_surface_scan} shows the  fidelity loss $\Delta W=1-W_2$ as a function of the velocity $v$ and the delay $T_0$ for typical experimental parameters given in Table~\ref{tab:experimental_params}. As intuitively expected, high velocity $v$ and small delay $T_0$ greatly enhance the fidelity of the logical qubit shuttling. For the shuttling velocity of 1--10~$\mathrm{m}/\mathrm{s}$ and the delays $T_0$ of 10--100~ns, the fidelity loss $\Delta W$ is of the order of  $10^{-5}$ for shuttling the qubit over 10~${\mu}$m.

\begin{figure}[tbp!]
\includegraphics[width=\columnwidth]{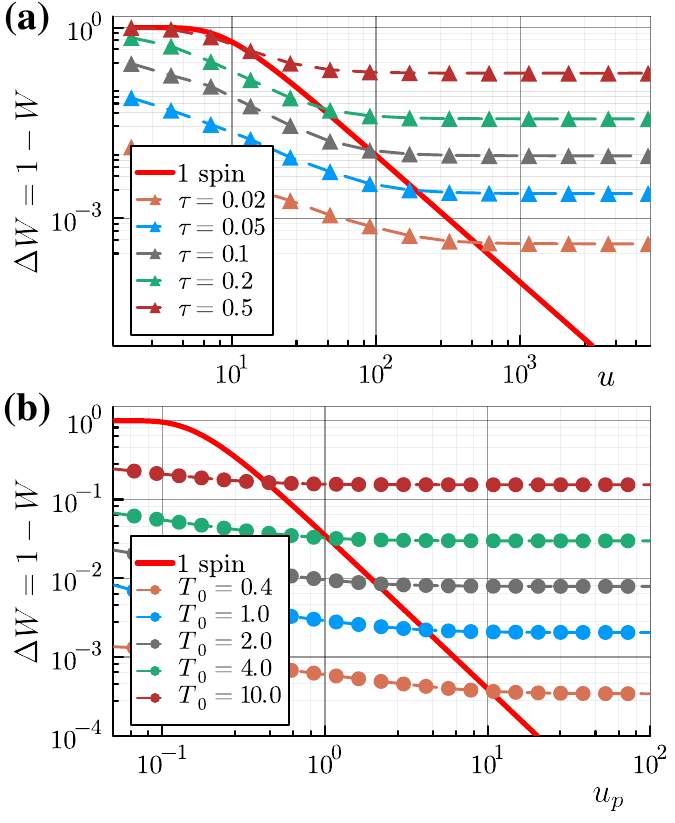}
\caption{
Fidelity loss $\Delta W$ of the shuttled ST qubit (dashed lines with symbols) as a function of the shuttling speed, compared with the fidelity loss of a single spin (solid red line). Different curves correspond to different delay times. 
{\bf (a)}: Shuttling through OU sheet, $\Delta W$ as a function of the dimensionless velocity $u=v\,\kappa_x/\kappa_t$ for different values of the delay $\tau=\kappa_t T_0$. The results correspond to $\gamma=100$ and $\sigma_{\mathrm B}=\sqrt{2}\,\kappa_t$, same as in Fig.~\ref{fig:F_surface_scan}.
{\bf (b)}: Shuttling through pink sheet, $\Delta W$ as a function of the normalized velocity $u_p=\kappa_x v$ for different values of the delay $T_0$. The results correspond to $\gamma=10$ and $\sigma_{\mathrm B}=0.07245$, with the cutoffs $\omega_2=10^3$ and $\omega_1=10^{-9}$.
With the dimensional parameters given in Table~\ref{tab:experimental_params}, the curves in both panels correspond to $T_{2,s}^*=20~\mu$s and the delays $T_0=0.4$, $1.0$, $2.0$, $4.0$, $10.0$ and $20\ \mu$s.}
\label{fig:F_single_vs_entangle}
\end{figure}

Shuttling of two entangled spins is feasible \cite{JadotMeunierEtalTwoSpinShuttling21}, but requires significant experimental  effort, and it is important to identify the situations where this effort is warranted. Comparison between the single-spin shuttling and shuttling of the ST  logical qubit is facilitated by introducing dimensionless quantities 
\begin{equation}
\eta =\kappa_t T_1,\;\; \tau = \kappa_t T_0,\;\; \gamma= \kappa_x L,\;\; u= v\, \kappa_x/\kappa_t,
\end{equation}
i.e.\ dimensionless shuttling time, length, delay and velocity, respectively.
For fast shuttling ($u\to\infty$), keeping the delay $\tau$ and the shuttling length $\gamma$ fixed, we have
\begin{equation}
\label{eq:stlargeu}
\chi_2\approx 2 \left(\sigma_\mathrm{B}/\kappa_t\right)^2 (1-\mathrm{e}^{-\gamma})\left(\tau + \mathrm{e}^{-\tau }-1\right)\ \ \mathrm{at}\ \ u\to \infty.
\end{equation}
In the opposite limit of slow shuttling ($u\to 0$) with fixed $\tau$ and $\gamma$, we obtain 
\begin{equation}
\label{eq:stsmallu}
\chi_2\approx 2 \left(\sigma_\mathrm{B}/\kappa_t\right)^2 \gamma \left(\tau +e^{-\tau}-1\right)\ \ \mathrm{at}\ \ u\to 0.
\end{equation}
In both limits, the dephasing factor $W_2=\exp{-\chi_2}$ approaches finite values, determined by the length $\gamma$ and the delay $\tau$. In contrast, for the single-spin shuttling, the dephasing factor $W_1=\exp{-\chi_1}$ quickly goes to zero at slow shuttling velocity,
\begin{equation}
\label{eq:w0smallu}
\chi_1\approx \left(\sigma_\mathrm{B}/\kappa_t\right)^2\, \gamma/u\ \ \mathrm{at}\ \ u\to 0,
\end{equation}
but approaches 1 for fast shuttling, since
\begin{equation}
\label{eq:w0largeu}
\chi_1\approx \left(\sigma_\mathrm{B}/\kappa_t\right)^2\,\left(\gamma+\mathrm{e}^{-\gamma}-1\right)/u^2\ \ \mathrm{at}\ \ u\to\infty.
\end{equation}
The comparison between the two modes of shuttling is illustrated in Fig.~\ref{fig:F_single_vs_entangle}(a). If the shuttling speed could be increased indefinitely then the single-spin encoding would always be more advantageous: the fidelity loss $\Delta W_1$ approaches zero at large $u$, while $\Delta W_2$ saturates at a finite value, determined by the delay $\tau$. 
However, shuttling at very high speed is not only difficult experimentally, but also extremely counter-productive due to non-adiabaticity and other detrimental effects mentioned above, and the two-spin shuttling is preferred for any finite velocity $u$ if the delay time is made small enough.

More generally, in the limit of short delays $\tau\to 0$,
\begin{equation}
\chi_2 \approx \sigma_\mathrm{B}^2\, T_0^2\,\left[\frac{\gamma}{1+u} \right. 
+ \left. \left(1-\mathrm{e}^{-\gamma\frac{1+u}{u}}\right)\frac{u^2}{(1+u)^2} \right],
\end{equation}
which for the experimentally interesting case of large-distance shuttling, $\gamma\gg 1$, becomes 
\begin{equation}
\label{eq:stsmalltau}
\chi_2\approx \left(\sigma_\mathrm{B}/\kappa_t\right)^2 \tau^2 \gamma /(1+u).
\end{equation}
Thus, even very sluggish shuttling of two entangled spins can achieve arbitrarily high fidelity over arbitrarily large distances as long as the delay time is short enough. Slow shuttling nullifies various errors associated with non-adiabaticity and suppresses the loss of fidelity caused by the interface roughness and spin-orbit coupling. Moreover, sufficiently slow shuttling becomes insensitive to the spatial variations of the quantization axis in Ge-based semiconductor structures \cite{WangVeldhorstEtalGeQDs24} if the shuttling is slow enough to be adiabatic with respect to such variations, thus eliminating yet another possible problem. Thus, the two-spin encoding (or more general 3- and 4-spin encodings of a logical qubit \cite{EOqubit2,
BurkardLaddPanNicholReviewQuDots23}) is a very promising approach to fault-tolerant shuttling-based semiconductor quantum processors.

The advantages of the ST encoding come from leveraging the noise correlations and are not limited to the specific model of OU sheet. For instance, similar analysis for a charge noise with $1/f$-type spectrum in time and exponentially decaying correlations in space (``pink'' random sheet) \cite{SupplementalMaterial} shows that the dephasing exponent for the two-spin shuttling through pink sheet is
\begin{equation}
\chi_{2,\rm p} \approx \left(\sigma_B^2 T_0 ^2/\Delta\right) \left[\gamma  \left(A - \ln{\tau_p} \right) + \ln{(u_p/\omega_1)}-1\right]
\end{equation}
for long shuttling length $\gamma=\kappa_x L \gg 1$ and short delay $\tau_p \equiv \kappa_x v\,T_0 = u_p T_0 \ll 1$, where $u_p=\kappa_x v$ is the normalized shuttling speed for pink sheet
\footnote{
In contrast with OU sheet, where the temporal fluctuations have a clearly defined timescale $\tau_c$, pink random sheet with its $1/f$ noise power spectrum does not have such a single well-defined timescale; therefore, the shuttling velocity is normalized differently for the two cases, and different notations are used, $u$ and $u_p$.}. 
Here $\Delta=\ln{(\omega_2/\omega_1)}$ is the normalization factor of the $1/f$ noise power spectrum, $\omega_2$ and $\omega_1$ are the high- and the low-frequency cutoffs, respectively, and the constant $A=(3/2)-\gamma_E\approx 0.923$ where $\gamma_E$ is Euler's constant.
Fig.~\ref{fig:F_single_vs_entangle}(b) shows the fidelity loss for a single spin and for two entangled spins shuttled through the pink sheet noise, exhibiting  the behavior qualitatively similar to the case of OU sheet.

Summarizing, in this work we studied dephasing during shuttling of two entangled spins. The spatiotemporal correlations of the decohering noise are of utmost importance in this regime, and we use realistic models to correctly take them into account. We assessed how the encoding of a logical qubit in a singlet-triplet (ST) subspace of two spins enhances the shuttling fidelity, and identified the regimes where such an encoding is particularly favorable. We have shown that even very slow shuttling of the ST qubit can maintain arbitrarily high fidelity over arbitrarily large distances as long as the delay time is short. We demonstrate that such encodings constitute a promising approach to reaching the fault tolerance threshold in shuttling-based semiconductor architectures.

We thank D.~P.~DiVincenzo, L.~M.~K.~Vandersypen, A.~Roitershtein, M.~Russ-Rimbach, M.~Veldhorst and X.~Hu for valuable discussions.
This work is part of the research programme NWO QuTech Physics Funding (QTECH, programme 172) with project number 16QTECH02, which is (partly) financed by the Dutch Research Council (NWO).
Research was partly sponsored by the Army Research Office and was accomplished under Award Number W911NF-23-1-0110. 
%
%
%
%
\bibliography{ref-v6}

\begin{thebibliography}{50}%
\makeatletter
\providecommand \@ifxundefined [1]{%
 \@ifx{#1\undefined}
}%
\providecommand \@ifnum [1]{%
 \ifnum #1\expandafter \@firstoftwo
 \else \expandafter \@secondoftwo
 \fi
}%
\providecommand \@ifx [1]{%
 \ifx #1\expandafter \@firstoftwo
 \else \expandafter \@secondoftwo
 \fi
}%
\providecommand \natexlab [1]{#1}%
\providecommand \enquote  [1]{``#1''}%
\providecommand \bibnamefont  [1]{#1}%
\providecommand \bibfnamefont [1]{#1}%
\providecommand \citenamefont [1]{#1}%
\providecommand \href@noop [0]{\@secondoftwo}%
\providecommand \href [0]{\begingroup \@sanitize@url \@href}%
\providecommand \@href[1]{\@@startlink{#1}\@@href}%
\providecommand \@@href[1]{\endgroup#1\@@endlink}%
\providecommand \@sanitize@url [0]{\catcode `\\12\catcode `\$12\catcode
  `\&12\catcode `\#12\catcode `\^12\catcode `\_12\catcode `\%12\relax}%
\providecommand \@@startlink[1]{}%
\providecommand \@@endlink[0]{}%
\providecommand \url  [0]{\begingroup\@sanitize@url \@url }%
\providecommand \@url [1]{\endgroup\@href {#1}{\urlprefix }}%
\providecommand \urlprefix  [0]{URL }%
\providecommand \Eprint [0]{\href }%
\providecommand \doibase [0]{https://doi.org/}%
\providecommand \selectlanguage [0]{\@gobble}%
\providecommand \bibinfo  [0]{\@secondoftwo}%
\providecommand \bibfield  [0]{\@secondoftwo}%
\providecommand \translation [1]{[#1]}%
\providecommand \BibitemOpen [0]{}%
\providecommand \bibitemStop [0]{}%
\providecommand \bibitemNoStop [0]{.\EOS\space}%
\providecommand \EOS [0]{\spacefactor3000\relax}%
\providecommand \BibitemShut  [1]{\csname bibitem#1\endcsname}%
\let\auto@bib@innerbib\@empty
\bibitem [{\citenamefont {Taylor}\ \emph {et~al.}(2005)\citenamefont {Taylor},
  \citenamefont {Engel}, \citenamefont {D{\"u}r}, \citenamefont {Yacoby},
  \citenamefont {Marcus}, \citenamefont {Zoller},\ and\ \citenamefont
  {Lukin}}]{FaultTolQuDotTaylorLukin05}%
  \BibitemOpen
  \bibfield  {author} {\bibinfo {author} {\bibfnamefont {J.~M.}\ \bibnamefont
  {Taylor}}, \bibinfo {author} {\bibfnamefont {H.-A.}\ \bibnamefont {Engel}},
  \bibinfo {author} {\bibfnamefont {W.}~\bibnamefont {D{\"u}r}}, \bibinfo
  {author} {\bibfnamefont {A.}~\bibnamefont {Yacoby}}, \bibinfo {author}
  {\bibfnamefont {C.~M.}\ \bibnamefont {Marcus}}, \bibinfo {author}
  {\bibfnamefont {P.}~\bibnamefont {Zoller}},\ and\ \bibinfo {author}
  {\bibfnamefont {M.~D.}\ \bibnamefont {Lukin}},\ }\bibfield  {title} {\bibinfo
  {title} {Fault-tolerant architecture for quantum computation using
  electrically controlled semiconductor spins},\ }\href
  {https://doi.org/https://doi.org/10.1038/nphys174} {\bibfield  {journal}
  {\bibinfo  {journal} {Nature Phys.}\ }\textbf {\bibinfo {volume} {1}},\
  \bibinfo {pages} {177} (\bibinfo {year} {2005})}\BibitemShut {NoStop}%
\bibitem [{\citenamefont {Hermelin}\ \emph {et~al.}(2011)\citenamefont
  {Hermelin}, \citenamefont {Takada}, \citenamefont {Yamamoto}, \citenamefont
  {Tarucha}, \citenamefont {Wieck}, \citenamefont {Saminadayar}, \citenamefont
  {B{\"a}uerle},\ and\ \citenamefont {Meunier}}]{ShuttlingSAWMeunier11}%
  \BibitemOpen
  \bibfield  {author} {\bibinfo {author} {\bibfnamefont {S.}~\bibnamefont
  {Hermelin}}, \bibinfo {author} {\bibfnamefont {S.}~\bibnamefont {Takada}},
  \bibinfo {author} {\bibfnamefont {M.}~\bibnamefont {Yamamoto}}, \bibinfo
  {author} {\bibfnamefont {S.}~\bibnamefont {Tarucha}}, \bibinfo {author}
  {\bibfnamefont {A.~D.}\ \bibnamefont {Wieck}}, \bibinfo {author}
  {\bibfnamefont {L.}~\bibnamefont {Saminadayar}}, \bibinfo {author}
  {\bibfnamefont {C.}~\bibnamefont {B{\"a}uerle}},\ and\ \bibinfo {author}
  {\bibfnamefont {T.}~\bibnamefont {Meunier}},\ }\bibfield  {title} {\bibinfo
  {title} {Electrons surfing on a sound wave as a platform for quantum optics
  with flying electrons},\ }\href
  {https://doi.org/https://doi.org/10.1038/nature10416} {\bibfield  {journal}
  {\bibinfo  {journal} {Nature}\ }\textbf {\bibinfo {volume} {477}},\ \bibinfo
  {pages} {435} (\bibinfo {year} {2011})}\BibitemShut {NoStop}%
\bibitem [{\citenamefont {Buonacorsi}\ \emph {et~al.}(2019)\citenamefont
  {Buonacorsi}, \citenamefont {Cai}, \citenamefont {Ramirez}, \citenamefont
  {Willick}, \citenamefont {Walker}, \citenamefont {Li}, \citenamefont {Shaw},
  \citenamefont {Xu}, \citenamefont {Benjamin},\ and\ \citenamefont
  {Baugh}}]{BenjaminBaughEtalTopoQCSi19}%
  \BibitemOpen
  \bibfield  {author} {\bibinfo {author} {\bibfnamefont {B.}~\bibnamefont
  {Buonacorsi}}, \bibinfo {author} {\bibfnamefont {Z.}~\bibnamefont {Cai}},
  \bibinfo {author} {\bibfnamefont {E.~B.}\ \bibnamefont {Ramirez}}, \bibinfo
  {author} {\bibfnamefont {K.~S.}\ \bibnamefont {Willick}}, \bibinfo {author}
  {\bibfnamefont {S.~M.}\ \bibnamefont {Walker}}, \bibinfo {author}
  {\bibfnamefont {J.}~\bibnamefont {Li}}, \bibinfo {author} {\bibfnamefont
  {B.~D.}\ \bibnamefont {Shaw}}, \bibinfo {author} {\bibfnamefont
  {X.}~\bibnamefont {Xu}}, \bibinfo {author} {\bibfnamefont {S.~C.}\
  \bibnamefont {Benjamin}},\ and\ \bibinfo {author} {\bibfnamefont
  {J.}~\bibnamefont {Baugh}},\ }\bibfield  {title} {\bibinfo {title} {Network
  architecture for a topological quantum computer in silicon},\ }\href
  {https://doi.org/10.1088/2058-9565/aaf3c4} {\bibfield  {journal} {\bibinfo
  {journal} {Quantum Sci. Technol.}\ }\textbf {\bibinfo {volume} {4}},\
  \bibinfo {pages} {025003} (\bibinfo {year} {2019})}\BibitemShut {NoStop}%
\bibitem [{\citenamefont {Vandersypen}\ \emph {et~al.}(2017)\citenamefont
  {Vandersypen}, \citenamefont {Bluhm}, \citenamefont {Clarke}, \citenamefont
  {Dzurak}, \citenamefont {Ishihara}, \citenamefont {Morello}, \citenamefont
  {Reilly}, \citenamefont {Schreiber},\ and\ \citenamefont
  {Veldhorst}}]{VdSBluhmVeldhorstEtalVision17}%
  \BibitemOpen
  \bibfield  {author} {\bibinfo {author} {\bibfnamefont {L.~M.~K.}\
  \bibnamefont {Vandersypen}}, \bibinfo {author} {\bibfnamefont
  {H.}~\bibnamefont {Bluhm}}, \bibinfo {author} {\bibfnamefont {J.~S.}\
  \bibnamefont {Clarke}}, \bibinfo {author} {\bibfnamefont {A.~S.}\
  \bibnamefont {Dzurak}}, \bibinfo {author} {\bibfnamefont {R.}~\bibnamefont
  {Ishihara}}, \bibinfo {author} {\bibfnamefont {A.}~\bibnamefont {Morello}},
  \bibinfo {author} {\bibfnamefont {D.~J.}\ \bibnamefont {Reilly}}, \bibinfo
  {author} {\bibfnamefont {L.~R.}\ \bibnamefont {Schreiber}},\ and\ \bibinfo
  {author} {\bibfnamefont {M.}~\bibnamefont {Veldhorst}},\ }\bibfield  {title}
  {\bibinfo {title} {Interfacing spin qubits in quantum dots and donors ---
  hot, dense, and coherent},\ }\href
  {https://doi.org/https://doi.org/10.1038/s41534-017-0038-y} {\bibfield
  {journal} {\bibinfo  {journal} {npj Quantum Inf.}\ }\textbf {\bibinfo
  {volume} {3}},\ \bibinfo {pages} {34} (\bibinfo {year} {2017})}\BibitemShut
  {NoStop}%
\bibitem [{\citenamefont {Langrock}\ \emph {et~al.}(2023)\citenamefont
  {Langrock}, \citenamefont {Krzywda}, \citenamefont {Focke}, \citenamefont
  {Seidler}, \citenamefont {Schreiber},\ and\ \citenamefont
  {Cywi\ifmmode~\acute{n}\else \'{n}\fi{}ski}}]{langrock_blueprint_2023}%
  \BibitemOpen
  \bibfield  {author} {\bibinfo {author} {\bibfnamefont {V.}~\bibnamefont
  {Langrock}}, \bibinfo {author} {\bibfnamefont {J.~A.}\ \bibnamefont
  {Krzywda}}, \bibinfo {author} {\bibfnamefont {N.}~\bibnamefont {Focke}},
  \bibinfo {author} {\bibfnamefont {I.}~\bibnamefont {Seidler}}, \bibinfo
  {author} {\bibfnamefont {L.~R.}\ \bibnamefont {Schreiber}},\ and\ \bibinfo
  {author} {\bibfnamefont {L.}~\bibnamefont {Cywi\ifmmode~\acute{n}\else
  \'{n}\fi{}ski}},\ }\bibfield  {title} {\bibinfo {title} {Blueprint of a
  scalable spin qubit shuttle device for coherent mid-range qubit transfer in
  disordered {Si/SiGe/SiO$_2$}},\ }\href
  {https://doi.org/10.1103/PRXQuantum.4.020305} {\bibfield  {journal} {\bibinfo
   {journal} {PRX Quantum}\ }\textbf {\bibinfo {volume} {4}},\ \bibinfo {pages}
  {020305} (\bibinfo {year} {2023})}\BibitemShut {NoStop}%
\bibitem [{\citenamefont {K{\"u}nne}\ \emph {et~al.}(2023)\citenamefont
  {K{\"u}nne}, \citenamefont {Willmes}, \citenamefont {Oberl{\"a}nder},
  \citenamefont {Gorjaew}, \citenamefont {Teske}, \citenamefont {Bhardwaj},
  \citenamefont {Beer}, \citenamefont {Kammerloher}, \citenamefont {Otten},
  \citenamefont {Seidler}, \citenamefont {Xue}, \citenamefont {Schreiber},\
  and\ \citenamefont {Bluhm}}]{KunneEtalSpinBusArch23}%
  \BibitemOpen
  \bibfield  {author} {\bibinfo {author} {\bibfnamefont {M.}~\bibnamefont
  {K{\"u}nne}}, \bibinfo {author} {\bibfnamefont {A.}~\bibnamefont {Willmes}},
  \bibinfo {author} {\bibfnamefont {M.}~\bibnamefont {Oberl{\"a}nder}},
  \bibinfo {author} {\bibfnamefont {C.}~\bibnamefont {Gorjaew}}, \bibinfo
  {author} {\bibfnamefont {J.~D.}\ \bibnamefont {Teske}}, \bibinfo {author}
  {\bibfnamefont {H.}~\bibnamefont {Bhardwaj}}, \bibinfo {author}
  {\bibfnamefont {M.}~\bibnamefont {Beer}}, \bibinfo {author} {\bibfnamefont
  {E.}~\bibnamefont {Kammerloher}}, \bibinfo {author} {\bibfnamefont
  {R.}~\bibnamefont {Otten}}, \bibinfo {author} {\bibfnamefont
  {I.}~\bibnamefont {Seidler}}, \bibinfo {author} {\bibfnamefont
  {R.}~\bibnamefont {Xue}}, \bibinfo {author} {\bibfnamefont {L.~R.}\
  \bibnamefont {Schreiber}},\ and\ \bibinfo {author} {\bibfnamefont
  {H.}~\bibnamefont {Bluhm}},\ }\href
  {https://doi.org/10.48550/arXiv.2306.16348} {\bibinfo {title} {The {SpinBus}
  architecture: Scaling spin qubits with electron shuttling}} (\bibinfo {year}
  {2023}),\ \Eprint {https://arxiv.org/abs/2306.16348} {arXiv:2306.16348}
  \BibitemShut {NoStop}%
\bibitem [{\citenamefont {Wang}\ \emph {et~al.}(2022)\citenamefont {Wang},
  \citenamefont {Ota}, \citenamefont {Edlbauer}, \citenamefont {Jadot},
  \citenamefont {Mortemousque}, \citenamefont {Richard}, \citenamefont
  {Okazaki}, \citenamefont {Nakamura}, \citenamefont {Ludwig}, \citenamefont
  {Wieck}, \citenamefont {Urdampilleta}, \citenamefont {Meunier}, \citenamefont
  {Kodera}, \citenamefont {Kaneko}, \citenamefont {Takada},\ and\ \citenamefont
  {B{\"a}uerle}}]{BauerleTakadaEtalAcousticTransportElectrons22}%
  \BibitemOpen
  \bibfield  {author} {\bibinfo {author} {\bibfnamefont {J.}~\bibnamefont
  {Wang}}, \bibinfo {author} {\bibfnamefont {S.}~\bibnamefont {Ota}}, \bibinfo
  {author} {\bibfnamefont {H.}~\bibnamefont {Edlbauer}}, \bibinfo {author}
  {\bibfnamefont {B.}~\bibnamefont {Jadot}}, \bibinfo {author} {\bibfnamefont
  {P.-A.}\ \bibnamefont {Mortemousque}}, \bibinfo {author} {\bibfnamefont
  {A.}~\bibnamefont {Richard}}, \bibinfo {author} {\bibfnamefont
  {Y.}~\bibnamefont {Okazaki}}, \bibinfo {author} {\bibfnamefont
  {S.}~\bibnamefont {Nakamura}}, \bibinfo {author} {\bibfnamefont
  {A.}~\bibnamefont {Ludwig}}, \bibinfo {author} {\bibfnamefont {A.~D.}\
  \bibnamefont {Wieck}}, \bibinfo {author} {\bibfnamefont {M.}~\bibnamefont
  {Urdampilleta}}, \bibinfo {author} {\bibfnamefont {T.}~\bibnamefont
  {Meunier}}, \bibinfo {author} {\bibfnamefont {T.}~\bibnamefont {Kodera}},
  \bibinfo {author} {\bibfnamefont {N.-H.}\ \bibnamefont {Kaneko}}, \bibinfo
  {author} {\bibfnamefont {S.}~\bibnamefont {Takada}},\ and\ \bibinfo {author}
  {\bibfnamefont {C.}~\bibnamefont {B{\"a}uerle}},\ }\bibfield  {title}
  {\bibinfo {title} {Generation of a single-cycle acoustic pulse: A scalable
  solution for transport in single-electron circuits},\ }\href
  {https://doi.org/10.1103/PhysRevX.12.031035} {\bibfield  {journal} {\bibinfo
  {journal} {Phys. Rev. X}\ }\textbf {\bibinfo {volume} {12}},\ \bibinfo
  {pages} {031035} (\bibinfo {year} {2022})}\BibitemShut {NoStop}%
\bibitem [{\citenamefont {Mills}\ \emph {et~al.}(2019)\citenamefont {Mills},
  \citenamefont {Zajac}, \citenamefont {Gullans}, \citenamefont {Schupp},
  \citenamefont {Hazard},\ and\ \citenamefont {Petta}}]{mills_shuttling_2019}%
  \BibitemOpen
  \bibfield  {author} {\bibinfo {author} {\bibfnamefont {A.~R.}\ \bibnamefont
  {Mills}}, \bibinfo {author} {\bibfnamefont {D.~M.}\ \bibnamefont {Zajac}},
  \bibinfo {author} {\bibfnamefont {M.~J.}\ \bibnamefont {Gullans}}, \bibinfo
  {author} {\bibfnamefont {F.~J.}\ \bibnamefont {Schupp}}, \bibinfo {author}
  {\bibfnamefont {T.~M.}\ \bibnamefont {Hazard}},\ and\ \bibinfo {author}
  {\bibfnamefont {J.~R.}\ \bibnamefont {Petta}},\ }\bibfield  {title} {\bibinfo
  {title} {Shuttling a single charge across a one-dimensional array of silicon
  quantum dots},\ }\href {https://doi.org/10.1038/s41467-019-08970-z}
  {\bibfield  {journal} {\bibinfo  {journal} {Nature Communications}\ }\textbf
  {\bibinfo {volume} {10}},\ \bibinfo {pages} {1063} (\bibinfo {year}
  {2019})}\BibitemShut {NoStop}%
\bibitem [{\citenamefont {Jnane}\ \emph {et~al.}(2022)\citenamefont {Jnane},
  \citenamefont {Undseth}, \citenamefont {Cai}, \citenamefont {Benjamin},\ and\
  \citenamefont {Koczor}}]{BenjaminEtalMulticoreQComp22}%
  \BibitemOpen
  \bibfield  {author} {\bibinfo {author} {\bibfnamefont {H.}~\bibnamefont
  {Jnane}}, \bibinfo {author} {\bibfnamefont {B.}~\bibnamefont {Undseth}},
  \bibinfo {author} {\bibfnamefont {Z.}~\bibnamefont {Cai}}, \bibinfo {author}
  {\bibfnamefont {S.~C.}\ \bibnamefont {Benjamin}},\ and\ \bibinfo {author}
  {\bibfnamefont {B.}~\bibnamefont {Koczor}},\ }\bibfield  {title} {\bibinfo
  {title} {Multicore quantum computing},\ }\href
  {https://doi.org/10.1103/PhysRevApplied.18.044064} {\bibfield  {journal}
  {\bibinfo  {journal} {Phys. Rev. Appl.}\ }\textbf {\bibinfo {volume} {18}},\
  \bibinfo {pages} {044064} (\bibinfo {year} {2022})}\BibitemShut {NoStop}%
\bibitem [{\citenamefont {Boter}\ \emph {et~al.}(2022)\citenamefont {Boter},
  \citenamefont {Dehollain}, \citenamefont {van Dijk}, \citenamefont {Xu},
  \citenamefont {Hensgens}, \citenamefont {Versluis}, \citenamefont {Naus},
  \citenamefont {Clarke}, \citenamefont {Veldhorst}, \citenamefont
  {Sebastiano},\ and\ \citenamefont
  {Vandersypen}}]{BoterVdsEtalSpiderWebArray22}%
  \BibitemOpen
  \bibfield  {author} {\bibinfo {author} {\bibfnamefont {J.~M.}\ \bibnamefont
  {Boter}}, \bibinfo {author} {\bibfnamefont {J.~P.}\ \bibnamefont
  {Dehollain}}, \bibinfo {author} {\bibfnamefont {J.~P.~G.}\ \bibnamefont {van
  Dijk}}, \bibinfo {author} {\bibfnamefont {Y.}~\bibnamefont {Xu}}, \bibinfo
  {author} {\bibfnamefont {T.}~\bibnamefont {Hensgens}}, \bibinfo {author}
  {\bibfnamefont {R.}~\bibnamefont {Versluis}}, \bibinfo {author}
  {\bibfnamefont {H.~W.~L.}\ \bibnamefont {Naus}}, \bibinfo {author}
  {\bibfnamefont {J.~S.}\ \bibnamefont {Clarke}}, \bibinfo {author}
  {\bibfnamefont {M.}~\bibnamefont {Veldhorst}}, \bibinfo {author}
  {\bibfnamefont {F.}~\bibnamefont {Sebastiano}},\ and\ \bibinfo {author}
  {\bibfnamefont {L.~M.~K.}\ \bibnamefont {Vandersypen}},\ }\bibfield  {title}
  {\bibinfo {title} {Spiderweb array: A sparse spin-qubit array},\ }\href
  {https://doi.org/10.1103/PhysRevApplied.18.024053} {\bibfield  {journal}
  {\bibinfo  {journal} {Phys. Rev. Appl.}\ }\textbf {\bibinfo {volume} {18}},\
  \bibinfo {pages} {024053} (\bibinfo {year} {2022})}\BibitemShut {NoStop}%
\bibitem [{\citenamefont {Noiri}\ \emph {et~al.}(2022)\citenamefont {Noiri},
  \citenamefont {Takeda}, \citenamefont {Nakajima}, \citenamefont {Kobayashi},
  \citenamefont {Sammak}, \citenamefont {Scappucci},\ and\ \citenamefont
  {Tarucha}}]{noiri_shuttling-based_2022}%
  \BibitemOpen
  \bibfield  {author} {\bibinfo {author} {\bibfnamefont {A.}~\bibnamefont
  {Noiri}}, \bibinfo {author} {\bibfnamefont {K.}~\bibnamefont {Takeda}},
  \bibinfo {author} {\bibfnamefont {T.}~\bibnamefont {Nakajima}}, \bibinfo
  {author} {\bibfnamefont {T.}~\bibnamefont {Kobayashi}}, \bibinfo {author}
  {\bibfnamefont {A.}~\bibnamefont {Sammak}}, \bibinfo {author} {\bibfnamefont
  {G.}~\bibnamefont {Scappucci}},\ and\ \bibinfo {author} {\bibfnamefont
  {S.}~\bibnamefont {Tarucha}},\ }\bibfield  {title} {\bibinfo {title} {A
  shuttling-based two-qubit logic gate for linking distant silicon quantum
  processors},\ }\href {https://doi.org/10.1038/s41467-022-33453-z} {\bibfield
  {journal} {\bibinfo  {journal} {Nature Communications}\ }\textbf {\bibinfo
  {volume} {13}},\ \bibinfo {pages} {5740} (\bibinfo {year}
  {2022})}\BibitemShut {NoStop}%
\bibitem [{\citenamefont {Wang}\ \emph {et~al.}(2024)\citenamefont {Wang},
  \citenamefont {John}, \citenamefont {Tidjani}, \citenamefont {Yu},
  \citenamefont {Ivlev}, \citenamefont {D{\'e}prez}, \citenamefont {van
  Riggelen-Doelman}, \citenamefont {Woods}, \citenamefont {Hendrickx},
  \citenamefont {Lawrie}, \citenamefont {Stehouwer}, \citenamefont
  {Oosterhout}, \citenamefont {Sammak}, \citenamefont {Friesen}, \citenamefont
  {Scappucci}, \citenamefont {de~Snoo}, \citenamefont {Rimbach-Russ},
  \citenamefont {Borsoi},\ and\ \citenamefont
  {Veldhorst}}]{WangVeldhorstEtalGeQDs24}%
  \BibitemOpen
  \bibfield  {author} {\bibinfo {author} {\bibfnamefont {C.-A.}\ \bibnamefont
  {Wang}}, \bibinfo {author} {\bibfnamefont {V.}~\bibnamefont {John}}, \bibinfo
  {author} {\bibfnamefont {H.}~\bibnamefont {Tidjani}}, \bibinfo {author}
  {\bibfnamefont {C.~X.}\ \bibnamefont {Yu}}, \bibinfo {author} {\bibfnamefont
  {A.~S.}\ \bibnamefont {Ivlev}}, \bibinfo {author} {\bibfnamefont
  {C.}~\bibnamefont {D{\'e}prez}}, \bibinfo {author} {\bibfnamefont
  {F.}~\bibnamefont {van Riggelen-Doelman}}, \bibinfo {author} {\bibfnamefont
  {B.~D.}\ \bibnamefont {Woods}}, \bibinfo {author} {\bibfnamefont {N.~W.}\
  \bibnamefont {Hendrickx}}, \bibinfo {author} {\bibfnamefont {W.~I.~L.}\
  \bibnamefont {Lawrie}}, \bibinfo {author} {\bibfnamefont {L.~E.~A.}\
  \bibnamefont {Stehouwer}}, \bibinfo {author} {\bibfnamefont {S.~D.}\
  \bibnamefont {Oosterhout}}, \bibinfo {author} {\bibfnamefont
  {A.}~\bibnamefont {Sammak}}, \bibinfo {author} {\bibfnamefont
  {M.}~\bibnamefont {Friesen}}, \bibinfo {author} {\bibfnamefont
  {G.}~\bibnamefont {Scappucci}}, \bibinfo {author} {\bibfnamefont {S.~L.}\
  \bibnamefont {de~Snoo}}, \bibinfo {author} {\bibfnamefont {M.}~\bibnamefont
  {Rimbach-Russ}}, \bibinfo {author} {\bibfnamefont {F.}~\bibnamefont
  {Borsoi}},\ and\ \bibinfo {author} {\bibfnamefont {M.}~\bibnamefont
  {Veldhorst}},\ }\bibfield  {title} {\bibinfo {title} {Operating semiconductor
  quantum processors with hopping spins},\ }\href
  {https://doi.org/10.1126/science.ado5915} {\bibfield  {journal} {\bibinfo
  {journal} {Science}\ }\textbf {\bibinfo {volume} {385}},\ \bibinfo {pages}
  {447} (\bibinfo {year} {2024})}\BibitemShut {NoStop}%
\bibitem [{\citenamefont {Seidler}\ \emph {et~al.}(2022)\citenamefont
  {Seidler}, \citenamefont {Struck}, \citenamefont {Xue}, \citenamefont
  {Focke}, \citenamefont {Trellenkamp}, \citenamefont {Bluhm},\ and\
  \citenamefont {Schreiber}}]{seidler_conveyor-mode_2022}%
  \BibitemOpen
  \bibfield  {author} {\bibinfo {author} {\bibfnamefont {I.}~\bibnamefont
  {Seidler}}, \bibinfo {author} {\bibfnamefont {T.}~\bibnamefont {Struck}},
  \bibinfo {author} {\bibfnamefont {R.}~\bibnamefont {Xue}}, \bibinfo {author}
  {\bibfnamefont {N.}~\bibnamefont {Focke}}, \bibinfo {author} {\bibfnamefont
  {S.}~\bibnamefont {Trellenkamp}}, \bibinfo {author} {\bibfnamefont
  {H.}~\bibnamefont {Bluhm}},\ and\ \bibinfo {author} {\bibfnamefont {L.~R.}\
  \bibnamefont {Schreiber}},\ }\bibfield  {title} {\bibinfo {title}
  {Conveyor-mode single-electron shuttling in {Si}/{SiGe} for a scalable
  quantum computing architecture},\ }\href
  {https://doi.org/10.1038/s41534-022-00615-2} {\bibfield  {journal} {\bibinfo
  {journal} {npj Quantum Information}\ }\textbf {\bibinfo {volume} {8}},\
  \bibinfo {pages} {100} (\bibinfo {year} {2022})}\BibitemShut {NoStop}%
\bibitem [{\citenamefont {Struck}\ \emph {et~al.}(2024)\citenamefont {Struck},
  \citenamefont {Volmer}, \citenamefont {Visser}, \citenamefont {Offermann},
  \citenamefont {Xue}, \citenamefont {Tu}, \citenamefont {Trellenkamp},
  \citenamefont {Cywi{\'n}ski}, \citenamefont {Bluhm},\ and\ \citenamefont
  {Schreiber}}]{StruckSchreiberEtalSpinPairShuttling23}%
  \BibitemOpen
  \bibfield  {author} {\bibinfo {author} {\bibfnamefont {T.}~\bibnamefont
  {Struck}}, \bibinfo {author} {\bibfnamefont {M.}~\bibnamefont {Volmer}},
  \bibinfo {author} {\bibfnamefont {L.}~\bibnamefont {Visser}}, \bibinfo
  {author} {\bibfnamefont {T.}~\bibnamefont {Offermann}}, \bibinfo {author}
  {\bibfnamefont {R.}~\bibnamefont {Xue}}, \bibinfo {author} {\bibfnamefont
  {J.-S.}\ \bibnamefont {Tu}}, \bibinfo {author} {\bibfnamefont
  {S.}~\bibnamefont {Trellenkamp}}, \bibinfo {author} {\bibfnamefont
  {{\L}.}~\bibnamefont {Cywi{\'n}ski}}, \bibinfo {author} {\bibfnamefont
  {H.}~\bibnamefont {Bluhm}},\ and\ \bibinfo {author} {\bibfnamefont {L.~R.}\
  \bibnamefont {Schreiber}},\ }\bibfield  {title} {\bibinfo {title}
  {Spin-{EPR}-pair separation by conveyor-mode single electron shuttling in
  {Si}/{SiGe}},\ }\href {https://doi.org/10.1038/s41467-024-45583-7} {\bibfield
   {journal} {\bibinfo  {journal} {Nature Communications}\ }\textbf {\bibinfo
  {volume} {15}},\ \bibinfo {pages} {1325} (\bibinfo {year}
  {2024})}\BibitemShut {NoStop}%
\bibitem [{\citenamefont {Smet}\ \emph {et~al.}(2024)\citenamefont {Smet},
  \citenamefont {Matsumoto}, \citenamefont {Zwerver}, \citenamefont {Tryputen},
  \citenamefont {de~Snoo}, \citenamefont {Amitonov}, \citenamefont {Sammak},
  \citenamefont {Samkharadze}, \citenamefont {G{\"u}l}, \citenamefont
  {Wasserman}, \citenamefont {Rimbach-Russ}, \citenamefont {Scappucci},\ and\
  \citenamefont {Vandersypen}}]{SmetVandersypenEtal24SpinShuttlSilicon}%
  \BibitemOpen
  \bibfield  {author} {\bibinfo {author} {\bibfnamefont {M.~D.}\ \bibnamefont
  {Smet}}, \bibinfo {author} {\bibfnamefont {Y.}~\bibnamefont {Matsumoto}},
  \bibinfo {author} {\bibfnamefont {A.-M.~J.}\ \bibnamefont {Zwerver}},
  \bibinfo {author} {\bibfnamefont {L.}~\bibnamefont {Tryputen}}, \bibinfo
  {author} {\bibfnamefont {S.~L.}\ \bibnamefont {de~Snoo}}, \bibinfo {author}
  {\bibfnamefont {S.~V.}\ \bibnamefont {Amitonov}}, \bibinfo {author}
  {\bibfnamefont {A.}~\bibnamefont {Sammak}}, \bibinfo {author} {\bibfnamefont
  {N.}~\bibnamefont {Samkharadze}}, \bibinfo {author} {\bibfnamefont
  {{\"O}.}~\bibnamefont {G{\"u}l}}, \bibinfo {author} {\bibfnamefont
  {R.~N.~M.}\ \bibnamefont {Wasserman}}, \bibinfo {author} {\bibfnamefont
  {M.}~\bibnamefont {Rimbach-Russ}}, \bibinfo {author} {\bibfnamefont
  {G.}~\bibnamefont {Scappucci}},\ and\ \bibinfo {author} {\bibfnamefont
  {L.~M.~K.}\ \bibnamefont {Vandersypen}},\ }\href
  {https://arxiv.org/abs/2406.07267} {\bibinfo {title} {High-fidelity
  single-spin shuttling in silicon}} (\bibinfo {year} {2024}),\ \Eprint
  {https://arxiv.org/abs/2406.07267} {arXiv:2406.07267 [cond-mat.mes-hall]}
  \BibitemShut {NoStop}%
\bibitem [{\citenamefont {Yoneda}\ \emph {et~al.}(2021)\citenamefont {Yoneda},
  \citenamefont {Huang}, \citenamefont {Feng}, \citenamefont {Yang},
  \citenamefont {Chan}, \citenamefont {Tanttu}, \citenamefont {Gilbert},
  \citenamefont {Leon}, \citenamefont {Hudson}, \citenamefont {Itoh},
  \citenamefont {Morello}, \citenamefont {Bartlett}, \citenamefont {Laucht},
  \citenamefont {Saraiva},\ and\ \citenamefont
  {Dzurak}}]{yoneda_coherent_2021}%
  \BibitemOpen
  \bibfield  {author} {\bibinfo {author} {\bibfnamefont {J.}~\bibnamefont
  {Yoneda}}, \bibinfo {author} {\bibfnamefont {W.}~\bibnamefont {Huang}},
  \bibinfo {author} {\bibfnamefont {M.}~\bibnamefont {Feng}}, \bibinfo {author}
  {\bibfnamefont {C.~H.}\ \bibnamefont {Yang}}, \bibinfo {author}
  {\bibfnamefont {K.~W.}\ \bibnamefont {Chan}}, \bibinfo {author}
  {\bibfnamefont {T.}~\bibnamefont {Tanttu}}, \bibinfo {author} {\bibfnamefont
  {W.}~\bibnamefont {Gilbert}}, \bibinfo {author} {\bibfnamefont {R.~C.~C.}\
  \bibnamefont {Leon}}, \bibinfo {author} {\bibfnamefont {F.~E.}\ \bibnamefont
  {Hudson}}, \bibinfo {author} {\bibfnamefont {K.~M.}\ \bibnamefont {Itoh}},
  \bibinfo {author} {\bibfnamefont {A.}~\bibnamefont {Morello}}, \bibinfo
  {author} {\bibfnamefont {S.~D.}\ \bibnamefont {Bartlett}}, \bibinfo {author}
  {\bibfnamefont {A.}~\bibnamefont {Laucht}}, \bibinfo {author} {\bibfnamefont
  {A.}~\bibnamefont {Saraiva}},\ and\ \bibinfo {author} {\bibfnamefont {A.~S.}\
  \bibnamefont {Dzurak}},\ }\bibfield  {title} {\bibinfo {title} {Coherent spin
  qubit transport in silicon},\ }\href
  {https://doi.org/10.1038/s41467-021-24371-7} {\bibfield  {journal} {\bibinfo
  {journal} {Nature Communications}\ }\textbf {\bibinfo {volume} {12}},\
  \bibinfo {pages} {4114} (\bibinfo {year} {2021})}\BibitemShut {NoStop}%
\bibitem [{\citenamefont {Volmer}\ \emph {et~al.}(2024)\citenamefont {Volmer},
  \citenamefont {Struck}, \citenamefont {Sala}, \citenamefont {Chen},
  \citenamefont {Oberl{\"a}nder}, \citenamefont {Offermann}, \citenamefont
  {Xue}, \citenamefont {Visser}, \citenamefont {Tu}, \citenamefont
  {Trellenkamp}, \citenamefont {Cywi{\'n}ski}, \citenamefont {Bluhm},\ and\
  \citenamefont {Schreiber}}]{VolmerStruckEtalValleySplit23}%
  \BibitemOpen
  \bibfield  {author} {\bibinfo {author} {\bibfnamefont {M.}~\bibnamefont
  {Volmer}}, \bibinfo {author} {\bibfnamefont {T.}~\bibnamefont {Struck}},
  \bibinfo {author} {\bibfnamefont {A.}~\bibnamefont {Sala}}, \bibinfo {author}
  {\bibfnamefont {B.}~\bibnamefont {Chen}}, \bibinfo {author} {\bibfnamefont
  {M.}~\bibnamefont {Oberl{\"a}nder}}, \bibinfo {author} {\bibfnamefont
  {T.}~\bibnamefont {Offermann}}, \bibinfo {author} {\bibfnamefont
  {R.}~\bibnamefont {Xue}}, \bibinfo {author} {\bibfnamefont {L.}~\bibnamefont
  {Visser}}, \bibinfo {author} {\bibfnamefont {J.-S.}\ \bibnamefont {Tu}},
  \bibinfo {author} {\bibfnamefont {S.}~\bibnamefont {Trellenkamp}}, \bibinfo
  {author} {\bibfnamefont {{\L}.}~\bibnamefont {Cywi{\'n}ski}}, \bibinfo
  {author} {\bibfnamefont {H.}~\bibnamefont {Bluhm}},\ and\ \bibinfo {author}
  {\bibfnamefont {L.~R.}\ \bibnamefont {Schreiber}},\ }\bibfield  {title}
  {\bibinfo {title} {Mapping of valley-splitting by conveyor-mode spin-coherent
  electron shuttling},\ }\href {https://doi.org/10.1038/s41534-024-00852-7}
  {\bibfield  {journal} {\bibinfo  {journal} {npj Quantum Inf.}\ }\textbf
  {\bibinfo {volume} {10}},\ \bibinfo {pages} {61} (\bibinfo {year}
  {2024})}\BibitemShut {NoStop}%
\bibitem [{\citenamefont {Jadot}\ \emph {et~al.}(2021)\citenamefont {Jadot},
  \citenamefont {Mortemousque}, \citenamefont {Chanrion}, \citenamefont
  {Thiney}, \citenamefont {Ludwig}, \citenamefont {Wieck}, \citenamefont
  {Urdampilleta}, \citenamefont {B{\"a}uerle},\ and\ \citenamefont
  {Meunier}}]{JadotMeunierEtalTwoSpinShuttling21}%
  \BibitemOpen
  \bibfield  {author} {\bibinfo {author} {\bibfnamefont {B.}~\bibnamefont
  {Jadot}}, \bibinfo {author} {\bibfnamefont {P.-A.}\ \bibnamefont
  {Mortemousque}}, \bibinfo {author} {\bibfnamefont {E.}~\bibnamefont
  {Chanrion}}, \bibinfo {author} {\bibfnamefont {V.}~\bibnamefont {Thiney}},
  \bibinfo {author} {\bibfnamefont {A.}~\bibnamefont {Ludwig}}, \bibinfo
  {author} {\bibfnamefont {A.~D.}\ \bibnamefont {Wieck}}, \bibinfo {author}
  {\bibfnamefont {M.}~\bibnamefont {Urdampilleta}}, \bibinfo {author}
  {\bibfnamefont {C.}~\bibnamefont {B{\"a}uerle}},\ and\ \bibinfo {author}
  {\bibfnamefont {T.}~\bibnamefont {Meunier}},\ }\bibfield  {title} {\bibinfo
  {title} {Distant spin entanglement via fast and coherent electron
  shuttling},\ }\href
  {https://doi.org/https://doi.org/10.1038/s41565-021-00846-y} {\bibfield
  {journal} {\bibinfo  {journal} {Nat. Nanotechnol.}\ }\textbf {\bibinfo
  {volume} {16}},\ \bibinfo {pages} {570} (\bibinfo {year} {2021})}\BibitemShut
  {NoStop}%
\bibitem [{\citenamefont {McNeil}\ \emph {et~al.}(2011)\citenamefont {McNeil},
  \citenamefont {Kataoka}, \citenamefont {Ford}, \citenamefont {Barnes},
  \citenamefont {Anderson}, \citenamefont {Jones}, \citenamefont {Farrer},\
  and\ \citenamefont {Ritchie}}]{ShuttlingSAWRitchie11}%
  \BibitemOpen
  \bibfield  {author} {\bibinfo {author} {\bibfnamefont {R.~P.~G.}\
  \bibnamefont {McNeil}}, \bibinfo {author} {\bibfnamefont {M.}~\bibnamefont
  {Kataoka}}, \bibinfo {author} {\bibfnamefont {C.~J.~B.}\ \bibnamefont
  {Ford}}, \bibinfo {author} {\bibfnamefont {C.~H.~W.}\ \bibnamefont {Barnes}},
  \bibinfo {author} {\bibfnamefont {D.}~\bibnamefont {Anderson}}, \bibinfo
  {author} {\bibfnamefont {G.~A.~C.}\ \bibnamefont {Jones}}, \bibinfo {author}
  {\bibfnamefont {I.}~\bibnamefont {Farrer}},\ and\ \bibinfo {author}
  {\bibfnamefont {D.~A.}\ \bibnamefont {Ritchie}},\ }\bibfield  {title}
  {\bibinfo {title} {On-demand single-electron transfer between distant quantum
  dots},\ }\href {https://doi.org/https://doi.org/10.1038/nature10444}
  {\bibfield  {journal} {\bibinfo  {journal} {Nature}\ }\textbf {\bibinfo
  {volume} {477}},\ \bibinfo {pages} {439} (\bibinfo {year}
  {2011})}\BibitemShut {NoStop}%
\bibitem [{\citenamefont {Huang}\ and\ \citenamefont
  {Hu}(2013)}]{HuangHuSpinRelaxShuttl13}%
  \BibitemOpen
  \bibfield  {author} {\bibinfo {author} {\bibfnamefont {P.}~\bibnamefont
  {Huang}}\ and\ \bibinfo {author} {\bibfnamefont {X.}~\bibnamefont {Hu}},\
  }\bibfield  {title} {\bibinfo {title} {Spin qubit relaxation in a moving
  quantum dot},\ }\href {https://doi.org/10.1103/PhysRevB.88.075301} {\bibfield
   {journal} {\bibinfo  {journal} {Phys. Rev. B}\ }\textbf {\bibinfo {volume}
  {88}},\ \bibinfo {pages} {075301} (\bibinfo {year} {2013})}\BibitemShut
  {NoStop}%
\bibitem [{\citenamefont {van Riggelen-Doelman}\ \emph
  {et~al.}(2023)\citenamefont {van Riggelen-Doelman}, \citenamefont {Wang},
  \citenamefont {de~Snoo}, \citenamefont {Lawrie}, \citenamefont {Hendrickx},
  \citenamefont {Rimbach-Russ}, \citenamefont {Sammak}, \citenamefont
  {Scappucci}, \citenamefont {D{\'e}prez},\ and\ \citenamefont
  {Veldhorst}}]{van_riggelen-doelman_coherent_2023}%
  \BibitemOpen
  \bibfield  {author} {\bibinfo {author} {\bibfnamefont {F.}~\bibnamefont {van
  Riggelen-Doelman}}, \bibinfo {author} {\bibfnamefont {C.-A.}\ \bibnamefont
  {Wang}}, \bibinfo {author} {\bibfnamefont {S.~L.}\ \bibnamefont {de~Snoo}},
  \bibinfo {author} {\bibfnamefont {W.~I.~L.}\ \bibnamefont {Lawrie}}, \bibinfo
  {author} {\bibfnamefont {N.~W.}\ \bibnamefont {Hendrickx}}, \bibinfo {author}
  {\bibfnamefont {M.}~\bibnamefont {Rimbach-Russ}}, \bibinfo {author}
  {\bibfnamefont {A.}~\bibnamefont {Sammak}}, \bibinfo {author} {\bibfnamefont
  {G.}~\bibnamefont {Scappucci}}, \bibinfo {author} {\bibfnamefont
  {C.}~\bibnamefont {D{\'e}prez}},\ and\ \bibinfo {author} {\bibfnamefont
  {M.}~\bibnamefont {Veldhorst}},\ }\href {http://arxiv.org/abs/2308.02406}
  {\bibinfo {title} {Coherent spin qubit shuttling through germanium quantum
  dots}} (\bibinfo {year} {2023}),\ \bibinfo {note} {arXiv:2308.02406
  [cond-mat, physics:quant-ph]}\BibitemShut {NoStop}%
\bibitem [{\citenamefont {Boter}\ \emph {et~al.}(2020)\citenamefont {Boter},
  \citenamefont {Xue}, \citenamefont {Kr\"ahenmann}, \citenamefont {Watson},
  \citenamefont {Premakumar}, \citenamefont {Ward}, \citenamefont {Savage},
  \citenamefont {Lagally}, \citenamefont {Friesen}, \citenamefont
  {Coppersmith}, \citenamefont {Eriksson}, \citenamefont {Joynt},\ and\
  \citenamefont {Vandersypen}}]{BoterJoyntVdSNoiseCorrBellStates20}%
  \BibitemOpen
  \bibfield  {author} {\bibinfo {author} {\bibfnamefont {J.~M.}\ \bibnamefont
  {Boter}}, \bibinfo {author} {\bibfnamefont {X.}~\bibnamefont {Xue}}, \bibinfo
  {author} {\bibfnamefont {T.}~\bibnamefont {Kr\"ahenmann}}, \bibinfo {author}
  {\bibfnamefont {T.~F.}\ \bibnamefont {Watson}}, \bibinfo {author}
  {\bibfnamefont {V.~N.}\ \bibnamefont {Premakumar}}, \bibinfo {author}
  {\bibfnamefont {D.~R.}\ \bibnamefont {Ward}}, \bibinfo {author}
  {\bibfnamefont {D.~E.}\ \bibnamefont {Savage}}, \bibinfo {author}
  {\bibfnamefont {M.~G.}\ \bibnamefont {Lagally}}, \bibinfo {author}
  {\bibfnamefont {M.}~\bibnamefont {Friesen}}, \bibinfo {author} {\bibfnamefont
  {S.~N.}\ \bibnamefont {Coppersmith}}, \bibinfo {author} {\bibfnamefont
  {M.~A.}\ \bibnamefont {Eriksson}}, \bibinfo {author} {\bibfnamefont
  {R.}~\bibnamefont {Joynt}},\ and\ \bibinfo {author} {\bibfnamefont
  {L.~M.~K.}\ \bibnamefont {Vandersypen}},\ }\bibfield  {title} {\bibinfo
  {title} {Spatial noise correlations in a {Si/SiGe} two-qubit device from
  {Bell} state coherences},\ }\href
  {https://doi.org/10.1103/PhysRevB.101.235133} {\bibfield  {journal} {\bibinfo
   {journal} {Phys. Rev. B}\ }\textbf {\bibinfo {volume} {101}},\ \bibinfo
  {pages} {235133} (\bibinfo {year} {2020})}\BibitemShut {NoStop}%
\bibitem [{\citenamefont {Mortemousque}\ \emph {et~al.}(2021)\citenamefont
  {Mortemousque}, \citenamefont {Jadot}, \citenamefont {Chanrion},
  \citenamefont {Thiney}, \citenamefont {B\"auerle}, \citenamefont {Ludwig},
  \citenamefont {Wieck}, \citenamefont {Urdampilleta},\ and\ \citenamefont
  {Meunier}}]{MortemousqueMeunierEtalShuttl2DArray21}%
  \BibitemOpen
  \bibfield  {author} {\bibinfo {author} {\bibfnamefont {P.-A.}\ \bibnamefont
  {Mortemousque}}, \bibinfo {author} {\bibfnamefont {B.}~\bibnamefont {Jadot}},
  \bibinfo {author} {\bibfnamefont {E.}~\bibnamefont {Chanrion}}, \bibinfo
  {author} {\bibfnamefont {V.}~\bibnamefont {Thiney}}, \bibinfo {author}
  {\bibfnamefont {C.}~\bibnamefont {B\"auerle}}, \bibinfo {author}
  {\bibfnamefont {A.}~\bibnamefont {Ludwig}}, \bibinfo {author} {\bibfnamefont
  {A.~D.}\ \bibnamefont {Wieck}}, \bibinfo {author} {\bibfnamefont
  {M.}~\bibnamefont {Urdampilleta}},\ and\ \bibinfo {author} {\bibfnamefont
  {T.}~\bibnamefont {Meunier}},\ }\bibfield  {title} {\bibinfo {title}
  {Enhanced spin coherence while displacing electron in a two-dimensional array
  of quantum dots},\ }\href {https://doi.org/10.1103/PRXQuantum.2.030331}
  {\bibfield  {journal} {\bibinfo  {journal} {PRX Quantum}\ }\textbf {\bibinfo
  {volume} {2}},\ \bibinfo {pages} {030331} (\bibinfo {year}
  {2021})}\BibitemShut {NoStop}%
\bibitem [{\citenamefont {Bosco}\ \emph {et~al.}(2024)\citenamefont {Bosco},
  \citenamefont {Zou},\ and\ \citenamefont
  {Loss}}]{BoscoZouLossHighFidShuttlingSOI23}%
  \BibitemOpen
  \bibfield  {author} {\bibinfo {author} {\bibfnamefont {S.}~\bibnamefont
  {Bosco}}, \bibinfo {author} {\bibfnamefont {J.}~\bibnamefont {Zou}},\ and\
  \bibinfo {author} {\bibfnamefont {D.}~\bibnamefont {Loss}},\ }\bibfield
  {title} {\bibinfo {title} {High-fidelity spin qubit shuttling via large
  spin-orbit interactions},\ }\href
  {https://doi.org/10.1103/PRXQuantum.5.020353} {\bibfield  {journal} {\bibinfo
   {journal} {PRX Quantum}\ }\textbf {\bibinfo {volume} {5}},\ \bibinfo {pages}
  {020353} (\bibinfo {year} {2024})}\BibitemShut {NoStop}%
\bibitem [{\citenamefont {Losert}\ \emph {et~al.}(2023)\citenamefont {Losert},
  \citenamefont {Eriksson}, \citenamefont {Joynt}, \citenamefont {Rahman},
  \citenamefont {Scappucci}, \citenamefont {Coppersmith},\ and\ \citenamefont
  {Friesen}}]{LosertFriesenValleySplit23}%
  \BibitemOpen
  \bibfield  {author} {\bibinfo {author} {\bibfnamefont {M.~P.}\ \bibnamefont
  {Losert}}, \bibinfo {author} {\bibfnamefont {M.~A.}\ \bibnamefont
  {Eriksson}}, \bibinfo {author} {\bibfnamefont {R.}~\bibnamefont {Joynt}},
  \bibinfo {author} {\bibfnamefont {R.}~\bibnamefont {Rahman}}, \bibinfo
  {author} {\bibfnamefont {G.}~\bibnamefont {Scappucci}}, \bibinfo {author}
  {\bibfnamefont {S.~N.}\ \bibnamefont {Coppersmith}},\ and\ \bibinfo {author}
  {\bibfnamefont {M.}~\bibnamefont {Friesen}},\ }\bibfield  {title} {\bibinfo
  {title} {Practical strategies for enhancing the valley splitting in {Si/SiGe}
  quantum wells},\ }\href {https://doi.org/10.1103/PhysRevB.108.125405}
  {\bibfield  {journal} {\bibinfo  {journal} {Phys. Rev. B}\ }\textbf {\bibinfo
  {volume} {108}},\ \bibinfo {pages} {125405} (\bibinfo {year}
  {2023})}\BibitemShut {NoStop}%
\bibitem [{\citenamefont {Hao}\ \emph {et~al.}(2014)\citenamefont {Hao},
  \citenamefont {Ruskov}, \citenamefont {Xiao}, \citenamefont {Tahan},\ and\
  \citenamefont {Jiang}}]{HaoRuskovTahanEtal14}%
  \BibitemOpen
  \bibfield  {author} {\bibinfo {author} {\bibfnamefont {X.}~\bibnamefont
  {Hao}}, \bibinfo {author} {\bibfnamefont {R.}~\bibnamefont {Ruskov}},
  \bibinfo {author} {\bibfnamefont {M.}~\bibnamefont {Xiao}}, \bibinfo {author}
  {\bibfnamefont {C.}~\bibnamefont {Tahan}},\ and\ \bibinfo {author}
  {\bibfnamefont {H.}~\bibnamefont {Jiang}},\ }\bibfield  {title} {\bibinfo
  {title} {Electron spin resonance and spin-valley physics in a silicon double
  quantum dot},\ }\href {https://doi.org/10.1038/ncomms4860} {\bibfield
  {journal} {\bibinfo  {journal} {Nat. Commun.}\ }\textbf {\bibinfo {volume}
  {5}},\ \bibinfo {pages} {3860} (\bibinfo {year} {2014})}\BibitemShut
  {NoStop}%
\bibitem [{\citenamefont {Zanardi}\ and\ \citenamefont
  {Rasetti}(1997)}]{ZanardiRasettiDFS97}%
  \BibitemOpen
  \bibfield  {author} {\bibinfo {author} {\bibfnamefont {P.}~\bibnamefont
  {Zanardi}}\ and\ \bibinfo {author} {\bibfnamefont {M.}~\bibnamefont
  {Rasetti}},\ }\bibfield  {title} {\bibinfo {title} {Noiseless quantum
  codes},\ }\href {https://doi.org/10.1103/PhysRevLett.79.3306} {\bibfield
  {journal} {\bibinfo  {journal} {Phys. Rev. Lett.}\ }\textbf {\bibinfo
  {volume} {79}},\ \bibinfo {pages} {3306} (\bibinfo {year}
  {1997})}\BibitemShut {NoStop}%
\bibitem [{\citenamefont {Viola}\ \emph {et~al.}(2001)\citenamefont {Viola},
  \citenamefont {Fortunato}, \citenamefont {Pravia}, \citenamefont {Knill},
  \citenamefont {Laflamme},\ and\ \citenamefont
  {Cory}}]{ViolaCoryEtalDFSExp01}%
  \BibitemOpen
  \bibfield  {author} {\bibinfo {author} {\bibfnamefont {L.}~\bibnamefont
  {Viola}}, \bibinfo {author} {\bibfnamefont {E.~M.}\ \bibnamefont
  {Fortunato}}, \bibinfo {author} {\bibfnamefont {M.~A.}\ \bibnamefont
  {Pravia}}, \bibinfo {author} {\bibfnamefont {E.}~\bibnamefont {Knill}},
  \bibinfo {author} {\bibfnamefont {R.}~\bibnamefont {Laflamme}},\ and\
  \bibinfo {author} {\bibfnamefont {D.~G.}\ \bibnamefont {Cory}},\ }\bibfield
  {title} {\bibinfo {title} {Experimental realization of noiseless subsystems
  for quantum information processing},\ }\href
  {https://doi.org/10.1126/science.1064460} {\bibfield  {journal} {\bibinfo
  {journal} {Science}\ }\textbf {\bibinfo {volume} {293}},\ \bibinfo {pages}
  {2059} (\bibinfo {year} {2001})}\BibitemShut {NoStop}%
\bibitem [{\citenamefont {Burkard}\ \emph {et~al.}(2023)\citenamefont
  {Burkard}, \citenamefont {Ladd}, \citenamefont {Pan}, \citenamefont
  {Nichol},\ and\ \citenamefont {Petta}}]{BurkardLaddPanNicholReviewQuDots23}%
  \BibitemOpen
  \bibfield  {author} {\bibinfo {author} {\bibfnamefont {G.}~\bibnamefont
  {Burkard}}, \bibinfo {author} {\bibfnamefont {T.~D.}\ \bibnamefont {Ladd}},
  \bibinfo {author} {\bibfnamefont {A.}~\bibnamefont {Pan}}, \bibinfo {author}
  {\bibfnamefont {J.~M.}\ \bibnamefont {Nichol}},\ and\ \bibinfo {author}
  {\bibfnamefont {J.~R.}\ \bibnamefont {Petta}},\ }\bibfield  {title} {\bibinfo
  {title} {Semiconductor spin qubits},\ }\href
  {https://doi.org/10.1103/RevModPhys.95.025003} {\bibfield  {journal}
  {\bibinfo  {journal} {Rev. Mod. Phys.}\ }\textbf {\bibinfo {volume} {95}},\
  \bibinfo {pages} {025003} (\bibinfo {year} {2023})}\BibitemShut {NoStop}%
\bibitem [{\citenamefont {Struck}\ \emph {et~al.}(2020)\citenamefont {Struck},
  \citenamefont {Hollmann}, \citenamefont {Schauer}, \citenamefont {Fedorets},
  \citenamefont {Schmidbauer}, \citenamefont {Sawano}, \citenamefont {Riemann},
  \citenamefont {Abrosimov}, \citenamefont {Cywi{\'n}ski}, \citenamefont
  {Bougeard},\ and\ \citenamefont
  {Schreiber}}]{StruckCywinskiSchreiber20QubitNoise}%
  \BibitemOpen
  \bibfield  {author} {\bibinfo {author} {\bibfnamefont {T.}~\bibnamefont
  {Struck}}, \bibinfo {author} {\bibfnamefont {A.}~\bibnamefont {Hollmann}},
  \bibinfo {author} {\bibfnamefont {F.}~\bibnamefont {Schauer}}, \bibinfo
  {author} {\bibfnamefont {O.}~\bibnamefont {Fedorets}}, \bibinfo {author}
  {\bibfnamefont {A.}~\bibnamefont {Schmidbauer}}, \bibinfo {author}
  {\bibfnamefont {K.}~\bibnamefont {Sawano}}, \bibinfo {author} {\bibfnamefont
  {H.}~\bibnamefont {Riemann}}, \bibinfo {author} {\bibfnamefont {N.~V.}\
  \bibnamefont {Abrosimov}}, \bibinfo {author} {\bibfnamefont
  {{\L}.}~\bibnamefont {Cywi{\'n}ski}}, \bibinfo {author} {\bibfnamefont
  {D.}~\bibnamefont {Bougeard}},\ and\ \bibinfo {author} {\bibfnamefont
  {L.~R.}\ \bibnamefont {Schreiber}},\ }\bibfield  {title} {\bibinfo {title}
  {Low-frequency spin qubit energy splitting noise in highly purified
  {$^{28}$Si}/{SiGe}},\ }\href {https://doi.org/10.1038/s41534-020-0276-2}
  {\bibfield  {journal} {\bibinfo  {journal} {npj Quantum Inf.}\ }\textbf
  {\bibinfo {volume} {6}},\ \bibinfo {pages} {40} (\bibinfo {year}
  {2020})}\BibitemShut {NoStop}%
\bibitem [{\citenamefont {Kepa}\ \emph {et~al.}(2023)\citenamefont {Kepa},
  \citenamefont {Cywi{\'n}ski},\ and\ \citenamefont
  {Krzywda}}]{KepaCywinskiKrzywdaSpinNoise23}%
  \BibitemOpen
  \bibfield  {author} {\bibinfo {author} {\bibfnamefont {M.}~\bibnamefont
  {Kepa}}, \bibinfo {author} {\bibfnamefont {{\L}.}~\bibnamefont
  {Cywi{\'n}ski}},\ and\ \bibinfo {author} {\bibfnamefont {J.~A.}\ \bibnamefont
  {Krzywda}},\ }\bibfield  {title} {\bibinfo {title} {Correlations of spin
  splitting and orbital fluctuations due to 1/f charge noise in the {Si}/{SiGe}
  quantum dot},\ }\href {https://doi.org/10.1063/5.0156358} {\bibfield
  {journal} {\bibinfo  {journal} {Applied Physics Letters}\ }\textbf {\bibinfo
  {volume} {123}},\ \bibinfo {pages} {034003} (\bibinfo {year}
  {2023})}\BibitemShut {NoStop}%
\bibitem [{\citenamefont {Zou}\ \emph {et~al.}(2023)\citenamefont {Zou},
  \citenamefont {Bosco},\ and\ \citenamefont {Loss}}]{zou_spatially_2023}%
  \BibitemOpen
  \bibfield  {author} {\bibinfo {author} {\bibfnamefont {J.}~\bibnamefont
  {Zou}}, \bibinfo {author} {\bibfnamefont {S.}~\bibnamefont {Bosco}},\ and\
  \bibinfo {author} {\bibfnamefont {D.}~\bibnamefont {Loss}},\ }\href
  {http://arxiv.org/abs/2308.03054} {\bibinfo {title} {Spatially correlated
  classical and quantum noise in driven qubits: {The} good, the bad, and the
  ugly}} (\bibinfo {year} {2023})\BibitemShut {NoStop}%
\bibitem [{\citenamefont {Shalak}\ \emph {et~al.}(2023)\citenamefont {Shalak},
  \citenamefont {Delerue},\ and\ \citenamefont
  {Niquet}}]{ShalakDelerueNiquetChargeNoiseSiHole23}%
  \BibitemOpen
  \bibfield  {author} {\bibinfo {author} {\bibfnamefont {B.}~\bibnamefont
  {Shalak}}, \bibinfo {author} {\bibfnamefont {C.}~\bibnamefont {Delerue}},\
  and\ \bibinfo {author} {\bibfnamefont {Y.-M.}\ \bibnamefont {Niquet}},\
  }\bibfield  {title} {\bibinfo {title} {Modeling of spin decoherence in a {Si}
  hole qubit perturbed by a single charge fluctuator},\ }\href
  {https://doi.org/10.1103/PhysRevB.107.125415} {\bibfield  {journal} {\bibinfo
   {journal} {Phys. Rev. B}\ }\textbf {\bibinfo {volume} {107}},\ \bibinfo
  {pages} {125415} (\bibinfo {year} {2023})}\BibitemShut {NoStop}%
\bibitem [{\citenamefont {Spence}\ \emph {et~al.}(2022)\citenamefont {Spence},
  \citenamefont {Cardoso-Paz}, \citenamefont {Michal}, \citenamefont
  {Chanrion}, \citenamefont {Niegemann}, \citenamefont {Jadot}, \citenamefont
  {Mortemousque}, \citenamefont {Klemt}, \citenamefont {Thiney}, \citenamefont
  {Bertrand}, \citenamefont {Hutin}, \citenamefont {B{\"a}uerle}, \citenamefont
  {Balestro}, \citenamefont {Vinet}, \citenamefont {Niquet}, \citenamefont
  {Meunier},\ and\ \citenamefont
  {Urdampilleta}}]{SpenceNiquetMeunierEtalChargeNoise22}%
  \BibitemOpen
  \bibfield  {author} {\bibinfo {author} {\bibfnamefont {C.}~\bibnamefont
  {Spence}}, \bibinfo {author} {\bibfnamefont {B.}~\bibnamefont {Cardoso-Paz}},
  \bibinfo {author} {\bibfnamefont {V.}~\bibnamefont {Michal}}, \bibinfo
  {author} {\bibfnamefont {E.}~\bibnamefont {Chanrion}}, \bibinfo {author}
  {\bibfnamefont {D.~J.}\ \bibnamefont {Niegemann}}, \bibinfo {author}
  {\bibfnamefont {B.}~\bibnamefont {Jadot}}, \bibinfo {author} {\bibfnamefont
  {P.-A.}\ \bibnamefont {Mortemousque}}, \bibinfo {author} {\bibfnamefont
  {B.}~\bibnamefont {Klemt}}, \bibinfo {author} {\bibfnamefont
  {V.}~\bibnamefont {Thiney}}, \bibinfo {author} {\bibfnamefont
  {B.}~\bibnamefont {Bertrand}}, \bibinfo {author} {\bibfnamefont
  {L.}~\bibnamefont {Hutin}}, \bibinfo {author} {\bibfnamefont
  {C.}~\bibnamefont {B{\"a}uerle}}, \bibinfo {author} {\bibfnamefont
  {F.}~\bibnamefont {Balestro}}, \bibinfo {author} {\bibfnamefont
  {M.}~\bibnamefont {Vinet}}, \bibinfo {author} {\bibfnamefont {Y.-M.}\
  \bibnamefont {Niquet}}, \bibinfo {author} {\bibfnamefont {T.}~\bibnamefont
  {Meunier}},\ and\ \bibinfo {author} {\bibfnamefont {M.}~\bibnamefont
  {Urdampilleta}},\ }\href {https://doi.org/10.48550/arXiv.2209.01853}
  {\bibinfo {title} {Probing charge noise in few electron {CMOS} quantum dots}}
  (\bibinfo {year} {2022}),\ \Eprint {https://arxiv.org/abs/2209.01853}
  {arXiv:2209.01853 [cond-mat.mes-hall]} \BibitemShut {NoStop}%
\bibitem [{\citenamefont {Shehata}\ \emph {et~al.}(2023)\citenamefont
  {Shehata}, \citenamefont {Simion}, \citenamefont {Li}, \citenamefont
  {Mohiyaddin}, \citenamefont {Wan}, \citenamefont {Mongillo}, \citenamefont
  {Govoreanu}, \citenamefont {Radu}, \citenamefont {De~Greve},\ and\
  \citenamefont {Van~Dorpe}}]{ShehataVanDorpeEtalChargeNoiseQuDots23}%
  \BibitemOpen
  \bibfield  {author} {\bibinfo {author} {\bibfnamefont {M.~M. E.~K.}\
  \bibnamefont {Shehata}}, \bibinfo {author} {\bibfnamefont {G.}~\bibnamefont
  {Simion}}, \bibinfo {author} {\bibfnamefont {R.}~\bibnamefont {Li}}, \bibinfo
  {author} {\bibfnamefont {F.~A.}\ \bibnamefont {Mohiyaddin}}, \bibinfo
  {author} {\bibfnamefont {D.}~\bibnamefont {Wan}}, \bibinfo {author}
  {\bibfnamefont {M.}~\bibnamefont {Mongillo}}, \bibinfo {author}
  {\bibfnamefont {B.}~\bibnamefont {Govoreanu}}, \bibinfo {author}
  {\bibfnamefont {I.}~\bibnamefont {Radu}}, \bibinfo {author} {\bibfnamefont
  {K.}~\bibnamefont {De~Greve}},\ and\ \bibinfo {author} {\bibfnamefont
  {P.}~\bibnamefont {Van~Dorpe}},\ }\bibfield  {title} {\bibinfo {title}
  {Modeling semiconductor spin qubits and their charge noise environment for
  quantum gate fidelity estimation},\ }\href
  {https://doi.org/10.1103/PhysRevB.108.045305} {\bibfield  {journal} {\bibinfo
   {journal} {Phys. Rev. B}\ }\textbf {\bibinfo {volume} {108}},\ \bibinfo
  {pages} {045305} (\bibinfo {year} {2023})}\BibitemShut {NoStop}%
\bibitem [{Note1()}]{Note1}%
  \BibitemOpen
  \bibinfo {note} {We assume that the {\protect \em direction} of the effective
  magnetic field is almost uniform in space and time, such that the
  longitudinal relaxation can be neglected; this is a good approximation for
  Si-based structures. In GaAs systems the longitudinal relaxation due to
  spin-orbit coupling is important \cite {HuangHuSpinRelaxShuttl13}, while for
  Ge-based systems the spatial variation of the direction of the effective
  magnetic field should be taken into account \cite
  {van_riggelen-doelman_coherent_2023,
  WangVeldhorstEtalGeQDs24,BoscoZouLossHighFidShuttlingSOI23}, and the approach
  presented here needs some modifications.}\BibitemShut {Stop}%
\bibitem [{Sup()}]{SupplementalMaterial}%
  \BibitemOpen
  \href@noop {} {}\bibinfo {note} {Supplemental Material}\BibitemShut {NoStop}%
\bibitem [{\citenamefont {Chentsov}(1956)}]{Chentsov_1956}%
  \BibitemOpen
  \bibfield  {author} {\bibinfo {author} {\bibfnamefont {N.}~\bibnamefont
  {Chentsov}},\ }\bibfield  {title} {\bibinfo {title} {Wiener random fields
  depending on several parameters},\ }\href@noop {} {\bibfield  {journal}
  {\bibinfo  {journal} {Doklady Akademii Nauk SSSR}\ }\textbf {\bibinfo
  {volume} {106}},\ \bibinfo {pages} {607} (\bibinfo {year}
  {1956})}\BibitemShut {NoStop}%
\bibitem [{\citenamefont {Kitagava}(1951)}]{Kitagava_1951}%
  \BibitemOpen
  \bibfield  {author} {\bibinfo {author} {\bibfnamefont {T.}~\bibnamefont
  {Kitagava}},\ }\bibfield  {title} {\bibinfo {title} {Analysis of variance
  applied to function spaces},\ }\href@noop {} {\bibfield  {journal} {\bibinfo
  {journal} {Memoirs of the Faculty of Science, Kyushu University Series A}\
  }\textbf {\bibinfo {volume} {6}},\ \bibinfo {pages} {41} (\bibinfo {year}
  {1951})}\BibitemShut {NoStop}%
\bibitem [{\citenamefont {Adler}(1981)}]{AdlerBookGeomRandF}%
  \BibitemOpen
  \bibfield  {author} {\bibinfo {author} {\bibfnamefont {R.~J.}\ \bibnamefont
  {Adler}},\ }\href@noop {} {\emph {\bibinfo {title} {The geometry of random
  fields}}}\ (\bibinfo  {publisher} {John Wiley \& Sons Inc},\ \bibinfo
  {address} {Chichester},\ \bibinfo {year} {1981})\BibitemShut {NoStop}%
\bibitem [{\citenamefont {Mokeev}\ \emph {et~al.}(2024)\citenamefont {Mokeev},
  \citenamefont {Zhang},\ and\ \citenamefont
  {Dobrovitski}}]{MokeevZhangDobrovitski24}%
  \BibitemOpen
  \bibfield  {author} {\bibinfo {author} {\bibfnamefont {A.~S.}\ \bibnamefont
  {Mokeev}}, \bibinfo {author} {\bibfnamefont {Y.-N.}\ \bibnamefont {Zhang}},\
  and\ \bibinfo {author} {\bibfnamefont {V.~V.}\ \bibnamefont {Dobrovitski}},\
  }\href {https://doi.org/10.48550/arXiv.2409.04404} {\bibinfo {title}
  {Modeling of decoherence and fidelity enhancement during transport of
  entangled qubits}} (\bibinfo {year} {2024}),\ \Eprint
  {https://arxiv.org/abs/2409.04404} {arXiv:2409.04404} \BibitemShut {NoStop}%
\bibitem [{\citenamefont {Kubo}(1954)}]{Kubo1954}%
  \BibitemOpen
  \bibfield  {author} {\bibinfo {author} {\bibfnamefont {R.}~\bibnamefont
  {Kubo}},\ }\bibfield  {title} {\bibinfo {title} {Note on the stochastic
  theory of resonance absorption},\ }\href@noop {} {\bibfield  {journal}
  {\bibinfo  {journal} {Journal of the Physical Society of Japan}\ }\textbf
  {\bibinfo {volume} {9}},\ \bibinfo {pages} {935} (\bibinfo {year}
  {1954})}\BibitemShut {NoStop}%
\bibitem [{\citenamefont {Klauder}\ and\ \citenamefont
  {Anderson}(1962)}]{KlauderAnderson62}%
  \BibitemOpen
  \bibfield  {author} {\bibinfo {author} {\bibfnamefont {J.~R.}\ \bibnamefont
  {Klauder}}\ and\ \bibinfo {author} {\bibfnamefont {P.~W.}\ \bibnamefont
  {Anderson}},\ }\bibfield  {title} {\bibinfo {title} {Spectral diffusion decay
  in spin resonance experiments},\ }\href
  {https://doi.org/10.1103/PhysRev.125.912} {\bibfield  {journal} {\bibinfo
  {journal} {Phys. Rev.}\ }\textbf {\bibinfo {volume} {125}},\ \bibinfo {pages}
  {912} (\bibinfo {year} {1962})}\BibitemShut {NoStop}%
\bibitem [{\citenamefont {Chandrasekhar}(1943)}]{ChandrasekharRandomProc}%
  \BibitemOpen
  \bibfield  {author} {\bibinfo {author} {\bibfnamefont {S.}~\bibnamefont
  {Chandrasekhar}},\ }\bibfield  {title} {\bibinfo {title} {Stochastic problems
  in physics and astronomy},\ }\href {https://doi.org/10.1103/RevModPhys.15.1}
  {\bibfield  {journal} {\bibinfo  {journal} {Rev. Mod. Phys.}\ }\textbf
  {\bibinfo {volume} {15}},\ \bibinfo {pages} {1} (\bibinfo {year}
  {1943})}\BibitemShut {NoStop}%
\bibitem [{\citenamefont {Rojas-Arias}\ \emph {et~al.}(2023)\citenamefont
  {Rojas-Arias}, \citenamefont {Noiri}, \citenamefont {Stano}, \citenamefont
  {Nakajima}, \citenamefont {Yoneda}, \citenamefont {Takeda}, \citenamefont
  {Kobayashi}, \citenamefont {Sammak}, \citenamefont {Scappucci}, \citenamefont
  {Loss},\ and\ \citenamefont {Tarucha}}]{rojas-arias_spatial_2023}%
  \BibitemOpen
  \bibfield  {author} {\bibinfo {author} {\bibfnamefont {J.~S.}\ \bibnamefont
  {Rojas-Arias}}, \bibinfo {author} {\bibfnamefont {A.}~\bibnamefont {Noiri}},
  \bibinfo {author} {\bibfnamefont {P.}~\bibnamefont {Stano}}, \bibinfo
  {author} {\bibfnamefont {T.}~\bibnamefont {Nakajima}}, \bibinfo {author}
  {\bibfnamefont {J.}~\bibnamefont {Yoneda}}, \bibinfo {author} {\bibfnamefont
  {K.}~\bibnamefont {Takeda}}, \bibinfo {author} {\bibfnamefont
  {T.}~\bibnamefont {Kobayashi}}, \bibinfo {author} {\bibfnamefont
  {A.}~\bibnamefont {Sammak}}, \bibinfo {author} {\bibfnamefont
  {G.}~\bibnamefont {Scappucci}}, \bibinfo {author} {\bibfnamefont
  {D.}~\bibnamefont {Loss}},\ and\ \bibinfo {author} {\bibfnamefont
  {S.}~\bibnamefont {Tarucha}},\ }\bibfield  {title} {\bibinfo {title} {Spatial
  noise correlations beyond nearest neighbors in ${}^{28}\mathrm{Si}/${Si}-{Ge}
  spin qubits},\ }\href {https://doi.org/10.1103/PhysRevApplied.20.054024}
  {\bibfield  {journal} {\bibinfo  {journal} {Phys. Rev. Appl.}\ }\textbf
  {\bibinfo {volume} {20}},\ \bibinfo {pages} {054024} (\bibinfo {year}
  {2023})}\BibitemShut {NoStop}%
\bibitem [{\citenamefont {Yoneda}\ \emph {et~al.}(2023)\citenamefont {Yoneda},
  \citenamefont {Rojas-Arias}, \citenamefont {Stano}, \citenamefont {Takeda},
  \citenamefont {Noiri}, \citenamefont {Nakajima}, \citenamefont {Loss},\ and\
  \citenamefont {Tarucha}}]{yoneda_noise-correlation_2023}%
  \BibitemOpen
  \bibfield  {author} {\bibinfo {author} {\bibfnamefont {J.}~\bibnamefont
  {Yoneda}}, \bibinfo {author} {\bibfnamefont {J.~S.}\ \bibnamefont
  {Rojas-Arias}}, \bibinfo {author} {\bibfnamefont {P.}~\bibnamefont {Stano}},
  \bibinfo {author} {\bibfnamefont {K.}~\bibnamefont {Takeda}}, \bibinfo
  {author} {\bibfnamefont {A.}~\bibnamefont {Noiri}}, \bibinfo {author}
  {\bibfnamefont {T.}~\bibnamefont {Nakajima}}, \bibinfo {author}
  {\bibfnamefont {D.}~\bibnamefont {Loss}},\ and\ \bibinfo {author}
  {\bibfnamefont {S.}~\bibnamefont {Tarucha}},\ }\bibfield  {title} {\bibinfo
  {title} {Noise-correlation spectrum for a pair of spin qubits in silicon},\
  }\href {https://doi.org/10.1038/s41567-023-02238-6} {\bibfield  {journal}
  {\bibinfo  {journal} {Nat. Phys.}\ }\textbf {\bibinfo {volume} {19}},\
  \bibinfo {pages} {1793} (\bibinfo {year} {2023})}\BibitemShut {NoStop}%
\bibitem [{\citenamefont {Cywi\ifmmode~\acute{n}\else \'{n}\fi{}ski}\ \emph
  {et~al.}(2009)\citenamefont {Cywi\ifmmode~\acute{n}\else \'{n}\fi{}ski},
  \citenamefont {Witzel},\ and\ \citenamefont
  {Das~Sarma}}]{CywinskiWitzelDasSarmaDD09}%
  \BibitemOpen
  \bibfield  {author} {\bibinfo {author} {\bibfnamefont {L.}~\bibnamefont
  {Cywi\ifmmode~\acute{n}\else \'{n}\fi{}ski}}, \bibinfo {author}
  {\bibfnamefont {W.~M.}\ \bibnamefont {Witzel}},\ and\ \bibinfo {author}
  {\bibfnamefont {S.}~\bibnamefont {Das~Sarma}},\ }\bibfield  {title} {\bibinfo
  {title} {Pure quantum dephasing of a solid-state electron spin qubit in a
  large nuclear spin bath coupled by long-range hyperfine-mediated
  interactions},\ }\href {https://doi.org/10.1103/PhysRevB.79.245314}
  {\bibfield  {journal} {\bibinfo  {journal} {Phys. Rev. B}\ }\textbf {\bibinfo
  {volume} {79}},\ \bibinfo {pages} {245314} (\bibinfo {year}
  {2009})}\BibitemShut {NoStop}%
\bibitem [{\citenamefont {Dobrovitski}\ \emph {et~al.}(2009)\citenamefont
  {Dobrovitski}, \citenamefont {Feiguin}, \citenamefont {Hanson},\ and\
  \citenamefont {Awschalom}}]{DobrEtal09OUnoise}%
  \BibitemOpen
  \bibfield  {author} {\bibinfo {author} {\bibfnamefont {V.~V.}\ \bibnamefont
  {Dobrovitski}}, \bibinfo {author} {\bibfnamefont {A.~E.}\ \bibnamefont
  {Feiguin}}, \bibinfo {author} {\bibfnamefont {R.}~\bibnamefont {Hanson}},\
  and\ \bibinfo {author} {\bibfnamefont {D.~D.}\ \bibnamefont {Awschalom}},\
  }\bibfield  {title} {\bibinfo {title} {Decay of {Rabi} oscillations by
  dipolar-coupled dynamical spin environments},\ }\href
  {https://doi.org/10.1103/PhysRevLett.102.237601} {\bibfield  {journal}
  {\bibinfo  {journal} {Phys. Rev. Lett.}\ }\textbf {\bibinfo {volume} {102}},\
  \bibinfo {pages} {237601} (\bibinfo {year} {2009})}\BibitemShut {NoStop}%
\bibitem [{\citenamefont {DiVincenzo}\ \emph {et~al.}(2000)\citenamefont
  {DiVincenzo}, \citenamefont {Bacon}, \citenamefont {Kempe}, \citenamefont
  {Burkard},\ and\ \citenamefont {Whaley}}]{EOqubit2}%
  \BibitemOpen
  \bibfield  {author} {\bibinfo {author} {\bibfnamefont {D.~P.}\ \bibnamefont
  {DiVincenzo}}, \bibinfo {author} {\bibfnamefont {D.}~\bibnamefont {Bacon}},
  \bibinfo {author} {\bibfnamefont {J.}~\bibnamefont {Kempe}}, \bibinfo
  {author} {\bibfnamefont {G.}~\bibnamefont {Burkard}},\ and\ \bibinfo {author}
  {\bibfnamefont {K.~B.}\ \bibnamefont {Whaley}},\ }\bibfield  {title}
  {\bibinfo {title} {Universal quantum computation with the exchange
  interaction},\ }\href {https://www.nature.com/articles/35042541} {\bibfield
  {journal} {\bibinfo  {journal} {Nature}\ }\textbf {\bibinfo {volume} {408}},\
  \bibinfo {pages} {339} (\bibinfo {year} {2000})}\BibitemShut {NoStop}%
\bibitem [{Note2()}]{Note2}%
  \BibitemOpen
  \bibinfo {note} {In contrast with OU sheet, where the temporal fluctuations
  have a clearly defined timescale $\tau _c$, pink random sheet with its $1/f$
  noise power spectrum does not have such a single well-defined timescale;
  therefore, the shuttling velocity is normalized differently for the two
  cases, and different notations are used, $u$ and $u_p$.}\BibitemShut {Stop}%
\end{thebibliography}%
\end{document}